\documentclass[a4paper,fleqn,usenatbib,useAMS]{mn3e}

\RequirePackage[colorlinks,citecolor=blue,urlcolor=blue]{hyperref}

\usepackage{amsmath, amssymb}
\usepackage{algpseudocode}
\usepackage{algorithm}
\usepackage{graphicx}
\usepackage{subfigure}
\usepackage{lineno}
\usepackage{afterpage}
\usepackage{comment}
\usepackage{enumitem}
\usepackage{multirow}

\let\hat\widehat

\newcommand\norm[1]{\|#1\|}

\newcommand\h{{\sf h}}

\newcommand\N{{\sf N}}

\newcommand\zlow{{\sf z_{low}}}
\newcommand\NGC{{\sf N_{GC}}}

\newcommand\pmean{{\sf p_{mean}}}
\newcommand\prms{{\sf p_{rms}}}
\newcommand\Fdensity{{\sf F_{density}}}

\newcommand\ra{{\sf \alpha_{2000}}}
\newcommand\dec{{\sf \delta_{2000}}}
\newcommand\density{{\sf density}}
\newcommand\HH{{\sf H}}
\newcommand\vra{{\sf v_{ra}}}
\newcommand\vdec{{\sf v_{dec}}}

\newcommand\dF{{\sf dF}}
\newcommand\UM{{\sf UM}}

\usepackage{pbox}

\title[Cosmic Web Reconstruction through Density Ridges: Catalogue]{
Cosmic Web Reconstruction through 
Density Ridges: Catalogue}
\author[Yen-Chi Chen et al.]{Yen-Chi Chen,$^{1,3}$\thanks{E-mail:
yenchic@andrew.cmu.edu}
Shirley Ho,$^{2,3}$
Jon Brinkmann,$^{4}$
Peter E. Freeman,$^{1,3}$ \newauthor
Christopher R. Genovese,$^{1,3}$
Donald P. Schneider,$^{5,6}$
Larry Wasserman$^{1,3}$
\\
$^{1}$Department of Statistics, Carnegie Mellon University, Pittsburgh, PA 15213, USA\\
$^{2}$Department of Physics, Carnegie Mellon University, Pittsburgh, PA 15213, USA\\
$^{3}$McWilliams Center for Cosmology, Carnegie Mellon University, Pittsburgh, PA 15213, USA\\
$^{4}$Apache Point Observatory, Sunspot, NM 88349, USA\\
$^{5}$Department of Astronomy and Astrophysics, The Pennsylvania State University, University Park, PA 16802, USA\\
$^{6}$Institute for Gravitation and the Cosmos, The Pennsylvania State University, University Park, PA 16802, USA}
\begin{document}


\pagerange{\pageref{firstpage}--\pageref{lastpage}} \pubyear{2015}

\maketitle

\label{firstpage}

\begin{abstract}
We construct a catalogue for filaments using a novel approach called
SCMS (subspace constrained mean shift; \citealt{Ozertem2011, 2015arXiv150105303C}).
 SCMS is a gradient-based method
that detects filaments through density ridges (smooth curves
tracing high-density regions).
A great advantage of SCMS is its uncertainty measure, which allows
an evaluation of the errors for the detected filaments.
To detect filaments, we use data from the
Sloan Digital Sky Survey, which consist of three galaxy samples: 
the NYU main galaxy sample (MGS),
the LOWZ sample and the CMASS sample.
Each of the three dataset covers different redshift regions 
so that the combined sample allows detection of filaments
up to $z=0.7$.

Our filament catalogue consists of a sequence of
two-dimensional filament maps at different redshifts
that provide several useful statistics
on the evolution cosmic web.
To construct the maps, we select spectroscopically confirmed 
galaxies within $0.050<z<0.700$ and
partition them into $130$ bins.
For each bin, we ignore the redshift, 
treating the galaxy observations as a 2-D data and detect filaments using SCMS.
The filament catalogue consists of 130 individual 
2-D filament maps, and
each map comprises points on the detected filaments
that describe the filamentary structures
at a particular redshift.

We also apply our filament catalogue to investigate
galaxy luminosity and its relation with distance to filament.
Using a volume-limited sample,
we find strong evidence ($6.1\sigma - 12.3\sigma$) that 
galaxies close to filaments are generally brighter
than those at significant distance from filaments.

\end{abstract}


\section{Introduction}
Matter in the Universe tends to be distributed in a network-like
large-scale structure which 
is known as the cosmic web \citep{Bond1996}.
The existence of this filamentary structure
has been confirmed observationally
and can be reproduced
in N-body simulations 
\citep{Jenkins1998,Colberg2005,2005Natur.435..629S, 2006MNRAS.370..656D}.
The large-scale structure comprises four distinct objects:
overdense clusters, interconnected filaments,
widespread sheet-like walls and large empty voids.
In this paper, we focus on cosmic filaments.




The principal approach to study large-scale structure is
by constructing a catalogue. For galaxy clusters,
several catalogues have been created; 
see, e.g., Abell catalogue \citep{1989ApJS...70....1A},
redMaPPer \citep{2014ApJ...785..104R, 2014ApJ...783...80R}, 
XCS \citep{2013ApJ...765...67M}, 
MCMC \citep{2011A&A...534A.109P}, 
Mantz \citep{2010MNRAS.406.1773M}, and 
Planck ESZ \citep{2011A&A...536A...8P}.
However, there exists few catalogues for filaments 
(a recent example can be found in \citealt{2014MNRAS.438.3465T}).
There are three reasons why high-quality filament catalogues are needed.

First, catalogues for filaments provide a
reference to other types of large-scale structures.
It is known that galaxy clusters are connected
by filaments; filaments are mainly distributed
on cosmic sheets/walls
and surround large empty voids.
With a catalogue of filaments, 
it is easier to identify other large-scale structures. 

Second, filamentary structure at different redshifts
can be used to probe cosmological models.
In N-body simulations, we observe how a small fluctuation
in the initial density field, magnified by gravitational force over time,
weaves matter into a web-like structure 
\citep{Jenkins1998,Colberg2005,2005Natur.435..629S, 2006MNRAS.370..656D}.
The evolution of the cosmic web depends on the initial
condition of the early Universe.
Thus, how filaments change as a function of redshift
conveys information about dark matter and dark energy.

Finally, filament catalogues make it easier to study
properties of filaments and their interaction with
nearby galaxies.
For example, recent simulations have shown that 
galaxy intrinsic alignments and luminosity are impacted
by nearby filaments \citep{2015MNRAS.448.3391C}.
Orientations of filaments are also found to be correlated
with the shape, angular momentum and peculiar velocity of dark matter haloes
\citep{2007MNRAS.381...41H, 2007MNRAS.375..489H, 2009MNRAS.398.1742H,2008MNRAS.389.1127P, 2009ApJ...706..747Z, 2013ApJ...779..160Z, forero2014cosmic}.
Despite the abundance of results on simulation studies,
few measurements regarding galaxies aligned along filaments have been obtained 
(see \cite{Jones2009} and \cite{2015ApJ...800..112G} for some results
in spin alignment and luminosity).

In this paper, we present a catalogue for filaments
using a novel approach called SCMS
(subspace constrained mean shift, \citealt{Ozertem2011}) that
models filaments as ridges of the galaxy probability density function.
SCMS first estimates galaxy density fields, then
uses a gradient ascent method to detect ridges;
ridges are curve-like, smooth structures that characterize
high-density regions.
SCMS has two appealing properties.
First, SCMS consistently detects filaments in the sense that
the intersections between filaments
generally are populated by galaxy clusters
(See Section \ref{sec::int} and \citealt{2015arXiv150105303C}), 
as confirmed by other galaxy cluster detections 
\citep{2014ApJ...785..104R, 2014ApJ...783...80R}.
Second, SCMS is equipped with a statistically
consistent measure of uncertainty 
\citep{2014arXiv1406.5663C}
that allows an evaluation of the error for filament detection.

To construct the filament maps,
we use a combined galaxy sample from Sloan
Digital Sky Survey (SDSS; \citealt{2000AJ....120.1579Y, 2011AJ....142...72E}) that
consists of the three datasets: NYU MGS, LOWZ and CMASS samples.
We focus on redshifts between $z=0.050-0.700$
since the observed number density within this region is sufficiently high
to generate statistically meaningful results.
We first partition the Universe according to redshift into thin slices
of width $\Delta z = 0.005$,
then perform SCMS within each slice.
The above process yields a series of filament maps that characterize
the filamentary structure of the Universe at different redshifts.
The variation of filament maps at different redshifts provides
information about the evolution of the Universe that can be further used
to constrain cosmology.
We can construct filament maps for future photometric surveys 
(e.g. LSST) or spectroscopic surveys (e.g. WFIRST and Euclid)
by applying SCMS to these data.

To demonstrate the usefulness of our filament maps, 
we investigate the relationship between the luminosity of a galaxy
and its distance to nearby filaments.
We separate the three samples (NYU MGS, LOWZ and CMASS)
and compare galaxy luminosity versus its distance to filaments.
There is strong evidence that galaxies near
filaments tend to be brighter than those away from
filaments.

In this paper, we assume a WMAP 7 
$\Lambda$CDM cosmology with $H_0 = 70$, $\Omega_m = 0.274$, 
and $\Omega_\Lambda = 0.726$ \citep{2012MNRAS.427.3435A,2014MNRAS.439...83A}.

\begin{figure*}
\centering
\subfigure[]
	{
	\includegraphics[width=2 in, height = 2 in]{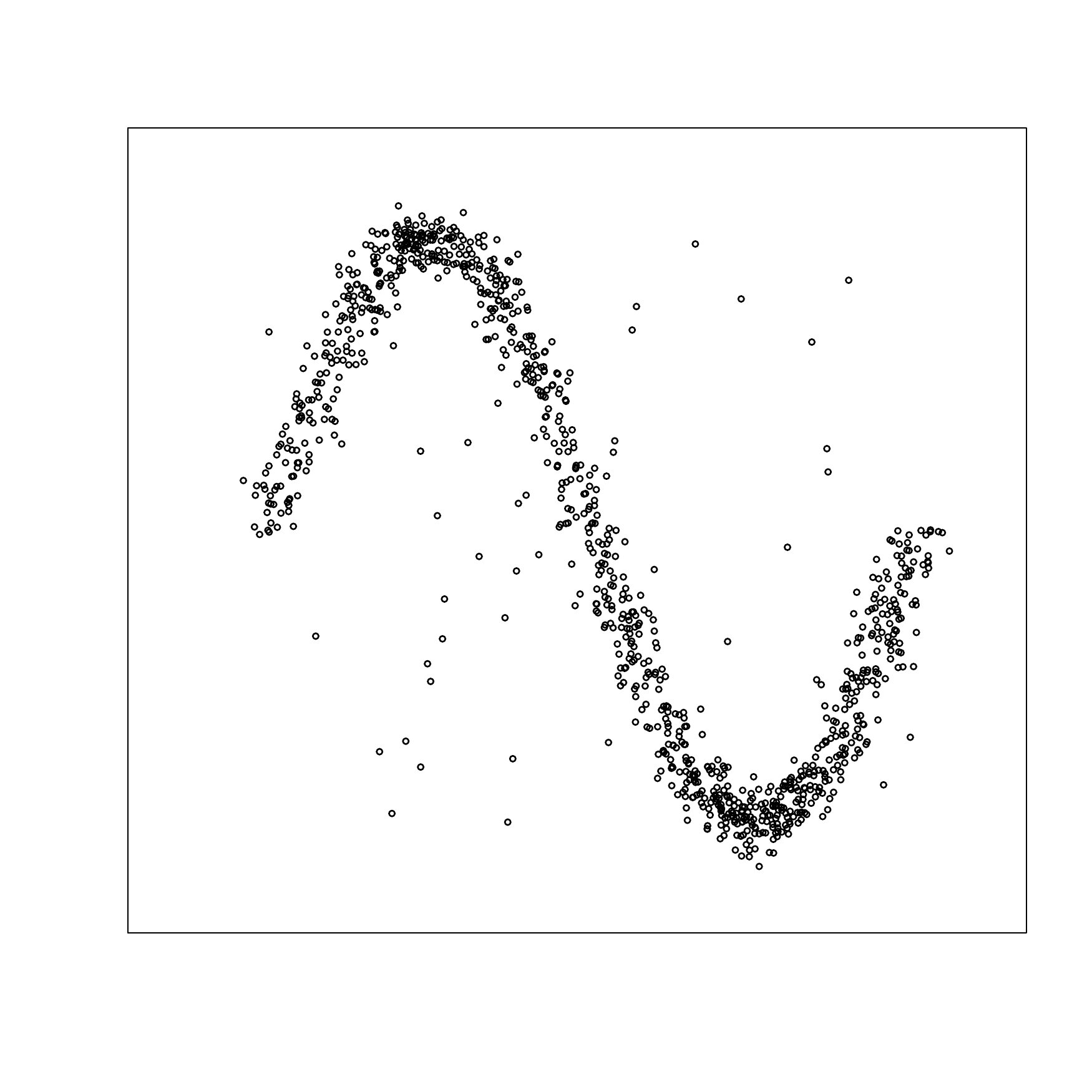}
	}
\subfigure[]
	{
	\includegraphics[width=2 in, height = 2 in]{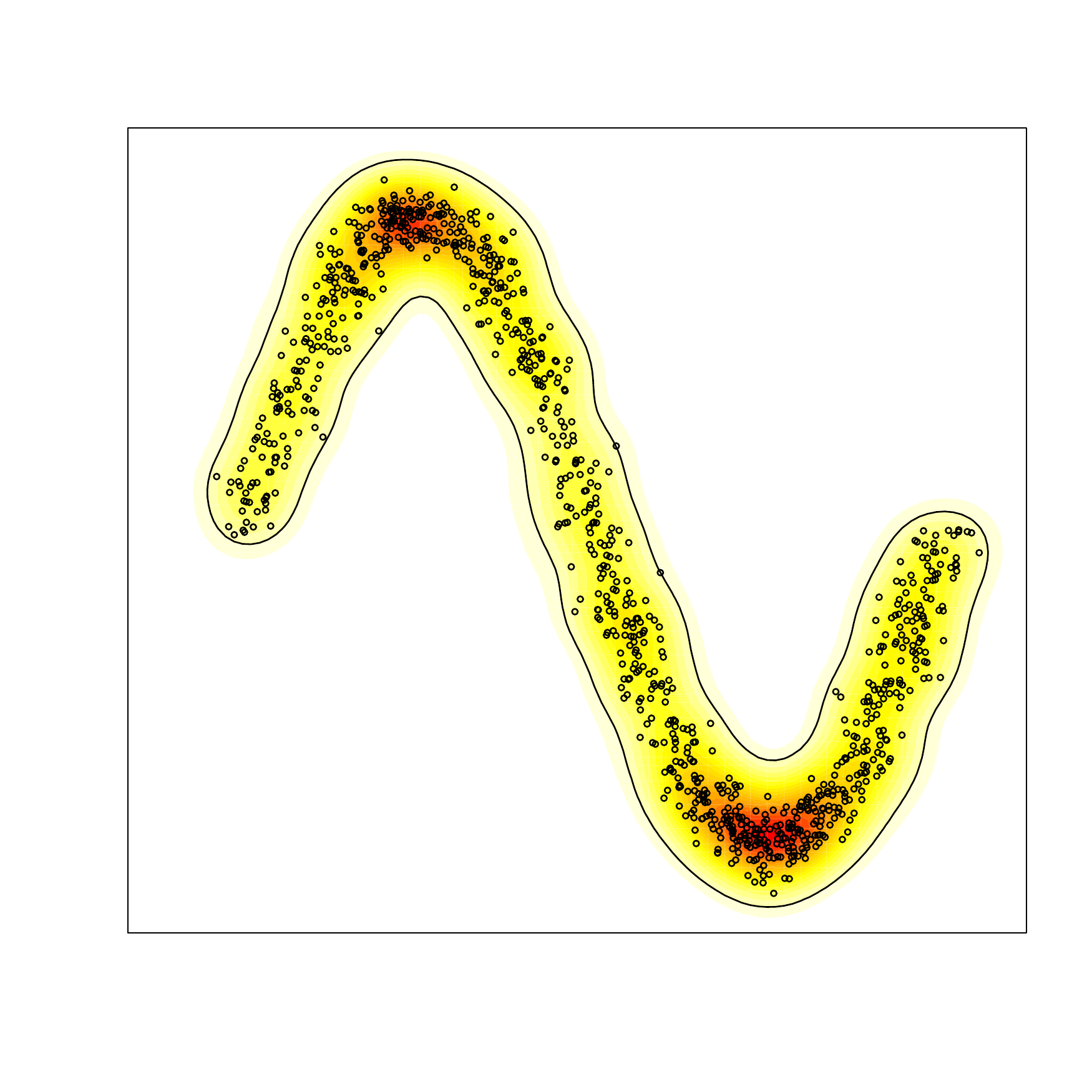}
	}
\subfigure[]
	{
	\includegraphics[width=2 in, height = 2 in]{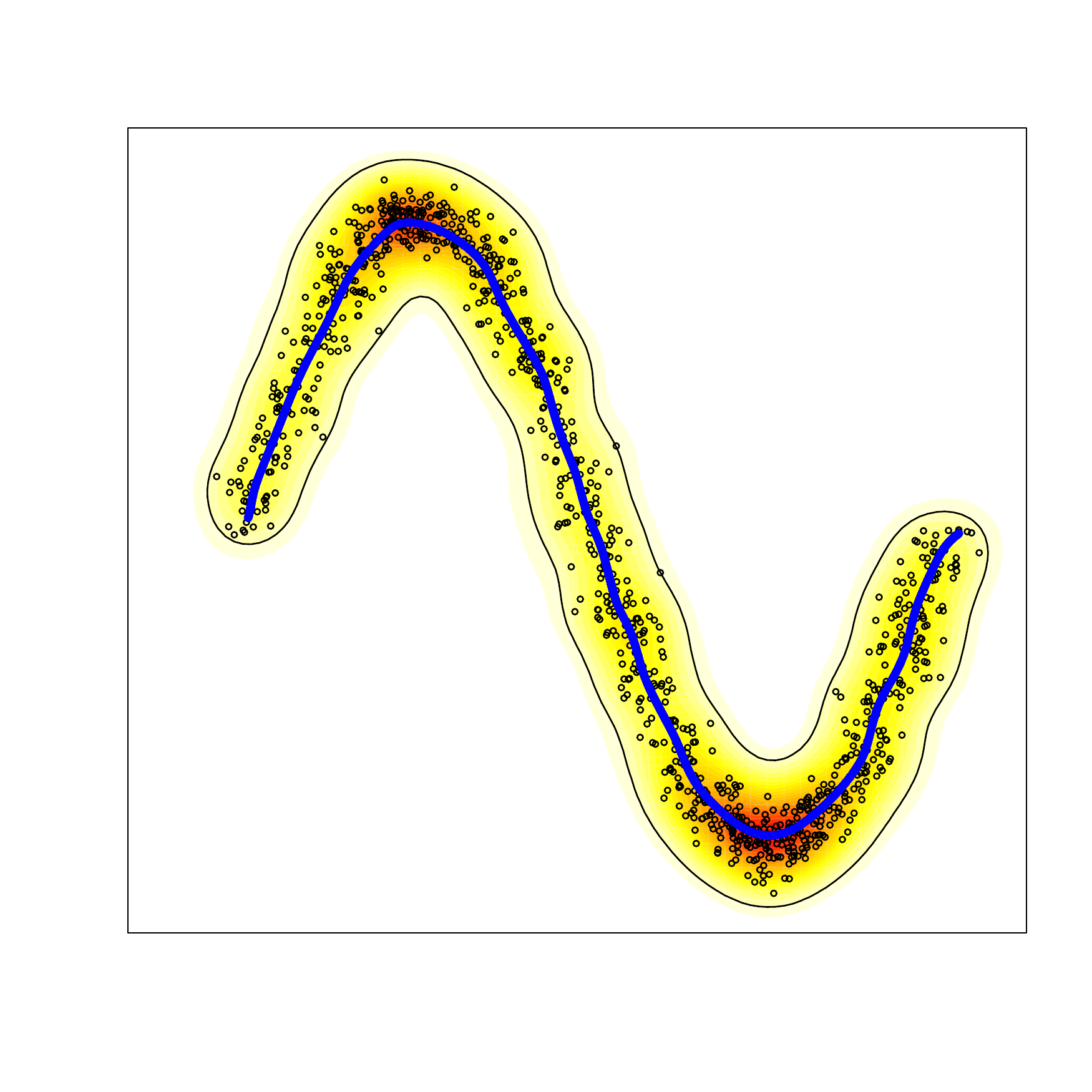}
	}
\caption{An illustration of the SCMS technique. {\bf (a)}: The original data. 
{\bf (b)}: Kernel density estimation (red-yellow: high density regions) 
and thresholding (removing points in low density regions).
 {\bf (c)}: The ridge estimation (blue curve).
Essentially, the SCMS technique is to identify ridges in
the galaxy density function estimated by the KDE; see
section {\S}\ref{sec::ridges} for more involved discussion.}
\label{fig:RidgeEstimation}
\end{figure*}

\section{Models and methods}	\label{sec::methods}

\subsection{SCMS: Detecting Filaments through Density Ridges} \label{sec::ridges}
We adopt the Subspace Constrain Mean Shift (SCMS; \citealt{Ozertem2011}) algorithm
to construct filament maps for
slices of our Universe at various redshifts.
We use the version of SCMS described in \cite{2015arXiv150105303C}.
SCMS detects filaments as galaxy density ridges \citep{2014arXiv1406.5663C}
and uses a three-step algorithm (density estimation, thresholding and gradient ascent)
to detect filaments.
Detailed implementations of SCMS can be found in 
\cite{Ozertem2011} and \cite{2015arXiv150105303C}.
Here we briefly discuss how density ridges are detected by SCMS.

Let $X_1,\cdots,X_n$ denote the locations of galaxies
and 
\begin{equation}
p(x) = \frac{1}{n\h^3} \sum_{i=1}^n K\left(\frac{\norm{x-X_i}}{\h}\right),
\label{eq::KDE}
\end{equation}
be the kernel density estimator where $h$ is
the smoothing bandwidth that controls the degree of smoothing,
$K$ is the Gaussian kernel, and $\norm{x-y}$
is the Euclidean distance between $x$ and $y$. 
Note that $p$ is also known as the kernel density estimator (KDE) in
statistical literature \citep{wasserman2006all}.
We further define $g(x) = \nabla p(x)$ and $H(x) = \nabla \nabla p(x)$
to be the gradient and the Hessian matrix of $p(x)$, respectively.

The \emph{density ridges }
\citep{Eberly1996,Ozertem2011,2012arXiv1212.5156G,2014arXiv1406.1803C,2014arXiv1406.5663C}
of $p$ are the collection of points
\begin{equation} \label{eq::ridge-def}
\begin{aligned}
R = \{x: v_j^T(x)g(x) = 0,j=2,3, 
\lambda_2(x) < 0\},
\end{aligned}
\end{equation}
where $v_j(x),\lambda_j(x)$ are the $j$-th eigenvector and eigenvalue, respectively, of $H(x)$
and $p_0$ is a density threshold.
Essentially, SCMS outputs a set of points on $R$. 
See Figure~\ref{fig:RidgeEstimation} for an example.

The idea of using eigen-structures of Hessian matrix
of the galaxy density
to detect filaments has been used
in other filament finders; see, e.g.,
the skeleton \citep{2006MNRAS.366.1201N}, 
the Multiscale Morphology Filter 
(MMF; \citealt{AragonCalvo2007,2010MNRAS.408.2163A}),
the Smoothed Hessian Major Axis Filament Finder 
(SHMAFF; \citealt{2010MNRAS.409..156B,2010MNRAS.406.1609B}),
the Spine method \citep{2010ApJ...723..364A},
and the DisPerSE model \citep{2011MNRAS.414..350S}.


\subsection{SCMS: Uncertainty Measure}	\label{sec::FU}
An appealing property of SCMS is its uncertainty measure.
We measure the error for filament detection via the bootstrap
technique \citep{Efron1979}.
We bootstrap the original data and apply SCMS to the bootstrap sample.
This exercise provides gives a set of bootstrap filaments.
We then compute the (projection) distances from $R$ (a filament detected
in the original sample) to each bootstrap filament.
Thus, each point on $R$ will be assigned an error value.
By repeating the bootstrap multiple times, for example $1,000$ times, 
every point on $R$ has $1,000$ error values.
The mean of these bootstrap error values for each point is the uncertainty measure
to the filaments detected by SCMS.
More detailed discussion on the uncertainty measure can be found in 
\cite{2015arXiv150105303C}.

\subsection{Slicing the Universe}
The observed galaxy locations contain three variables, namely,
the right ascension $\ra$, the declination $\dec$, and the redshift $z$.
We partition the range of redshift into several small intervals;
this procedure slices of the Universe.
We apply SCMS to galaxies within each slice
to detect filaments.

We slice the Universe for three purposes.
First, this action removes the Finger-of-God effect since galaxies at the same slice
share nearly the same redshift.
The Finger-of-God effect is produced by the small peculiar velocities of galaxies,
so galaxy clusters are stretched out along the line of sight
in redshift space.
Thus, most filaments appear to be pointing toward the Earth although
they may not really stretch along the line of sight.

Second, slicing the redshift Universe reduces the computational cost dramatically.
There are two barriers for computational complexity for SCMS.
One is the number of initialized points.
In the third step of SCMS (filament detection),
we must select many initial grid points to perform 
an ascending process. This ascending process, called 
subspace constrained mean shift,
pushes points until they arrive at ridges.
The size of the Universe greatly increases as the redshift increases,
so we need many grid points to
detect filaments.
The other barrier for the computation is that
SCMS requires evaluation of the Hessian of density function,
which is known to be computationally intensive if
the number of points is large or the dimension is high.
Taking slices of the Universe
reduces the dimension to two, and for each slice
the sample size (the number of galaxies) is small
so that SCMS can be performed within reasonable time.

Third, slicing the Universe according to the redshift allows a comparison of
filamentary structures
at different redshifts.
Characteristics of filaments at different redshifts
reveal information about the nature of our Universe that
can be used to constrain cosmological parameters.

\section{The SDSS Data}
We use a combined SDSS dataset that contains main galaxy sample (MGS)
from DR7 and LOWZ and CMASS samples from DR12 \citep{2015arXiv150100963A}.
We describe the datasets in detail in the following sections:

\subsection{The NYU MGS Catalogue}

The SDSS DR7 \citep{2009ApJS..182..543A}
contains the completed data set of SDSS-I and SDSS-II. These surveys 
obtained wide-field CCD photometry \citep{1998AJ....116.3040G,2006AJ....131.2332G} 
in five passbands 
(\emph{u, g, r, i, z} \citealt{1996AJ....111.1748F, 2010AJ....139.1628D}), 
internally calibrated using the `uber-calibration' 
process described in \cite{2008ApJ...674.1217P}, 
amassing a total footprint of 11,663 deg$^2$. 
From this imaging data, galaxies within a footprint of 9380 deg$^2$ \citep{2009ApJS..182..543A}
were selected for spectroscopic observation as part of the main galaxy sample 
\citep{2002AJ....124.1810S}, which, to good approximation, consists of all galaxies 
with $r_{pet} < 17.77$, where $r_{pet}$ is the extinction-corrected r-band 
Petrosian magnitude. In this analysis we do not consider the Luminous 
Red Galaxy extension of this program to higher redshift \citep{2001AJ....122.2267E}. 

We obtain the SDSS DR7 Main Galaxy Sample from the NYU value-added 
catalog (NYU VAGC, \citealt{2005AJ....129.2562B,2008ApJ...674.1217P,2008ApJS..175..297A}). 
It includes K-corrected absolute 
magnitudes, and detailed information on the mask. This sample uses galaxies 
with 14.5 $<r_{pet}<$ 17.6. The $r_{pet}>$ 14.5 limit ensures that only galaxies 
with reliable SDSS photometry are used and the $r_{pet} < 17.6$ allows a 
homogeneous selection over the full footprint of 6141 deg$^2$
\citep{2005AJ....129.2562B}. 
Galaxies that did not obtain a redshift due to fibre collisions are assigned 
the redshift of their nearest neighbour.

\subsection{The LOWZ and CMASS Catalogues}

The LOWZ and CMASS samples are from
from Data Release 12 \citep{2015arXiv150100963A}
of the Sloan Digital Sky Survey SDSS.
Together, SDSS I, II, and III 
imaged over one third of the sky
($14{,}555$ deg$^2$) in \emph{u, g, r, i, z} photometric bandpasses
to a limiting magnitude of $r\simeq 22.5$.
The imaging data were
processed through a series of pipelines that perform astrometric
calibration \citep{2003AJ....125.1559P}, photometric reduction \citep{2001ASPC..238..269L}, and
photometric calibration \citep{2008ApJ...674.1217P}. All of the imaging was reprocessed as part
of SDSS Data Release 8 \citep{2011ApJS..193...29A}.

The Baryon Oscillation Spectroscopic Survey (BOSS) of SDSS-III
has obtained spectra and redshifts for 1.35 million
galaxies over a footprint covering $10{,}000$ square
degrees. These galaxies are selected from the SDSS imaging \citep{2011ApJS..193...29A} and 
were observed together with $160{,}000$ quasars and approximately
$100{,}000$ ancillary targets.  The targets are assigned to tiles of
diameter $3^\circ$ using an algorithm \citep{2003AJ....125.2276B} that adopts to the
density of targets on the sky \citep{2003AJ....125.2276B}. Spectra are obtained
using the double-armed BOSS spectrographs \citep{2013AJ....146...32S}. 
Each observation is performed in a series of $900$-second exposures,
integrating until a minimum signal-to-noise ratio is achieved for the
faint galaxy targets. This approach ensures a homogeneous data set with a high
redshift completeness of more than $97$\% over the full survey
footprint. Redshifts are extracted from the spectra using the methods
described in \cite{2012AJ....144..144B}. A summary of the survey design appears
in \citet{2011AJ....142...72E}, a full description of BOSS is provided in \citet{2013AJ....145...10D}.

BOSS selects two classes of galaxies to be targeted for spectroscopy 
: `LOWZ' and `CMASS' (we refer the reader to \citealt{2014MNRAS.439...83A} for further description of these classes). 
For the LOWZ sample, 
the effective redshift is $z_{\rm eff} =0.32$, slightly lower than that of the
SDSS-II luminous red galaxies (LRGs)
as we place a redshift cut $z<0.43$. 
The CMASS selection yields a sample with a median redshift $z = 0.57$ and a
stellar mass that peaks at $\log_{10}(M/M_{\odot}) = 11.3$ \citep{2013MNRAS.435.2764M}.
Most CMASS targets are central galaxies residing in dark matter haloes of mass
$\sim10^{13}h^{-1}M_\odot$.

\section{Filament Maps}
\begin{figure*}
\centering
	\includegraphics[width=5 in]{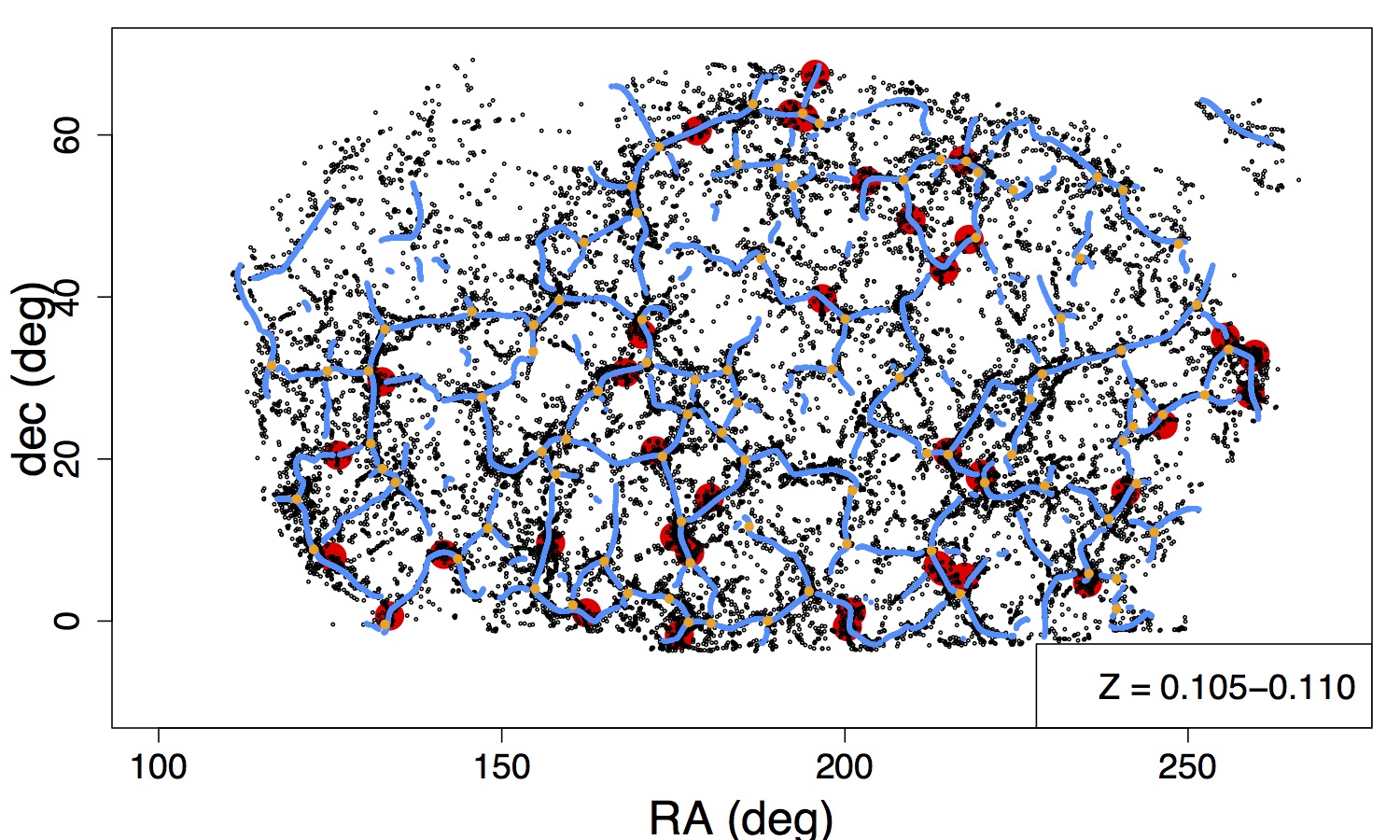}
	\includegraphics[width=5 in]{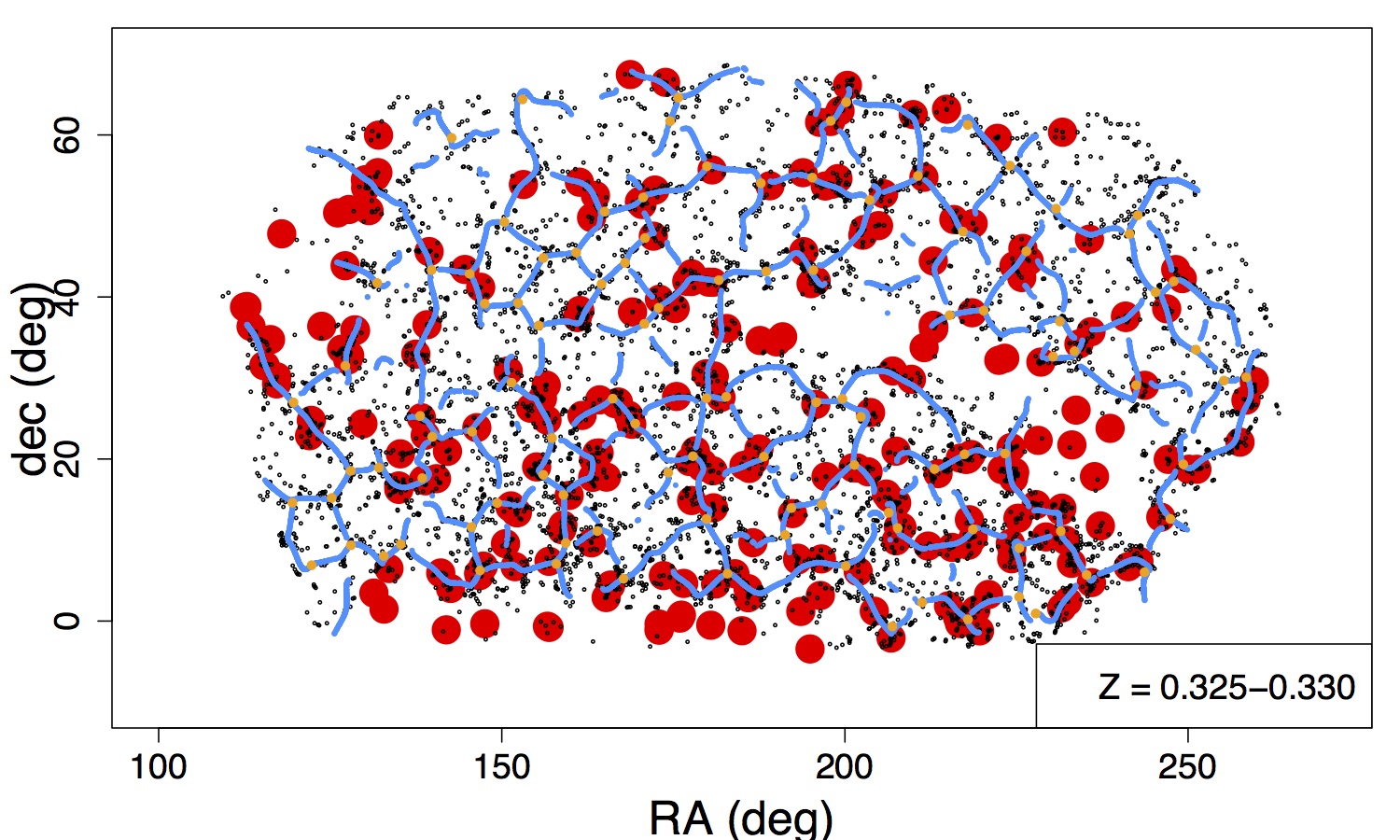}
	\includegraphics[width=5 in]{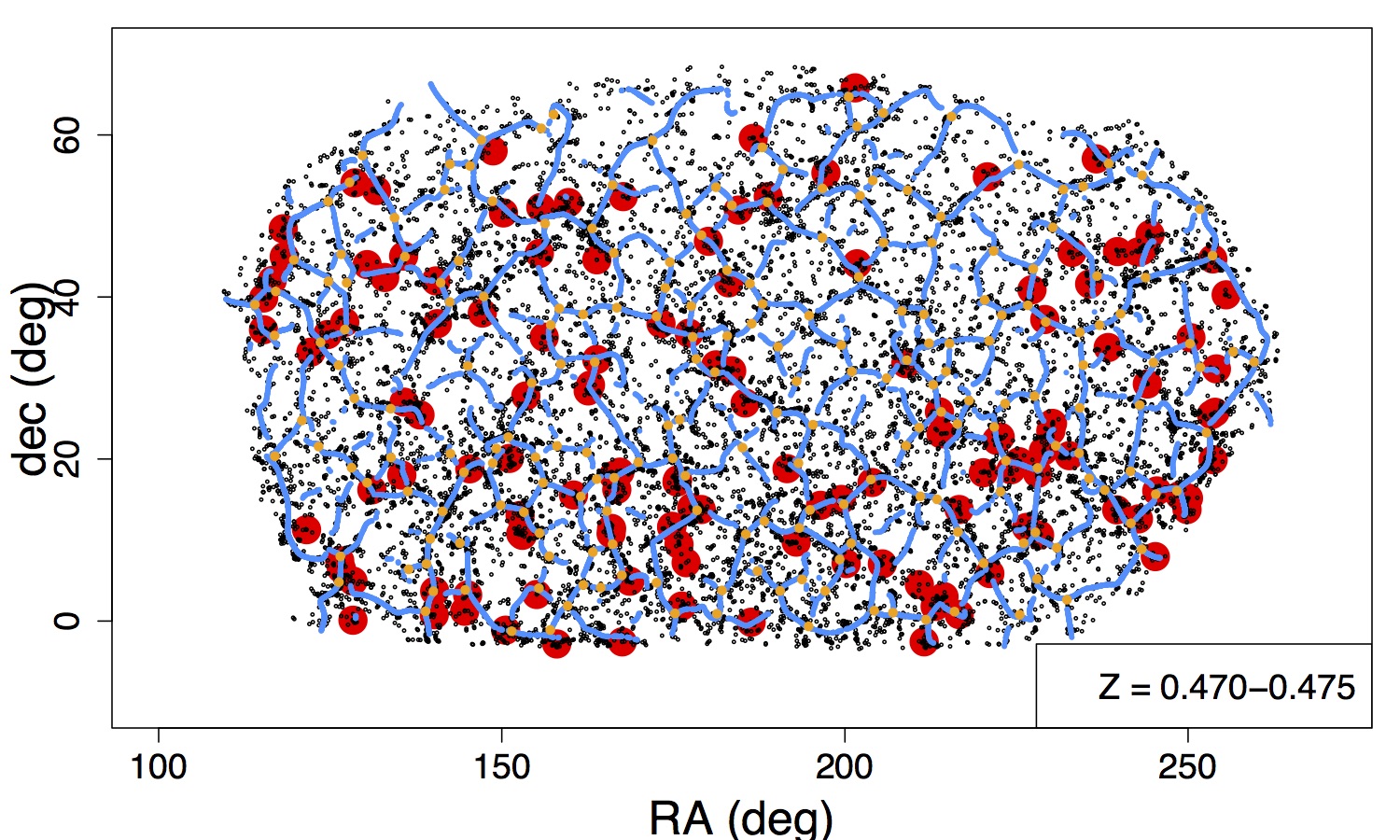}
\caption{Examples for filament maps from the SDSS data. 
From top to bottom: $z=0.105-0.110$ (NYU MGS), 
$0.325-0.330$ (LOWZ) and $0.470-0.475$ (CMASS). 
The blue curves are detected filaments,
the red dots are galaxy clusters from redMaPPer catalogue,
and
the orange dots are intersections for filaments (details can be found in 
Appendix \ref{sec::int}).}
\label{fig::FMex}
\end{figure*}

\subsection{Construction of Filament Maps}
We construct filament maps using the three galaxy catalogues: NYU MGS, LOWZ and CMASS.
Figure~\ref{fig::FMex} presents some examples of constructed filaments (blue) 
with galaxies (black) and galaxy clusters (red)
from the redMaPPer catalogue.
Our construction of filament maps consists of the following steps:
\begin{itemize}
\item[1.] Slice the sample between $0.050<z<0.700$ into $130$ slices
of width $\Delta z=0.005$ .
\item[2.] Within each slice, select galaxies within
\begin{equation*}
150\degr<\ra<200\degr,\quad 5\degr<\dec<30\degr
\end{equation*}
since this is a relatively complete region for all three 
galaxy catalogues.
\item[3.] Using KDE, compute the mean density and 
the root mean square (RMS) density for the 
selected galaxies.
\item[4.] Using the root mean square density as a threshold level in SCMS, construct filament maps.
The RMS density is used in SCMS as a threshold level to stabilize the algorithm.
\item[5.] Apply masks of galaxy catalogues to eliminate filaments outside the region of observations.
\item[6.] At each point on filaments, compute the filament's local direction.
\end{itemize}

\subsection{Filament Maps}

\begin{figure}
\centering
	\includegraphics[height=2.5 in, width=2.5 in]{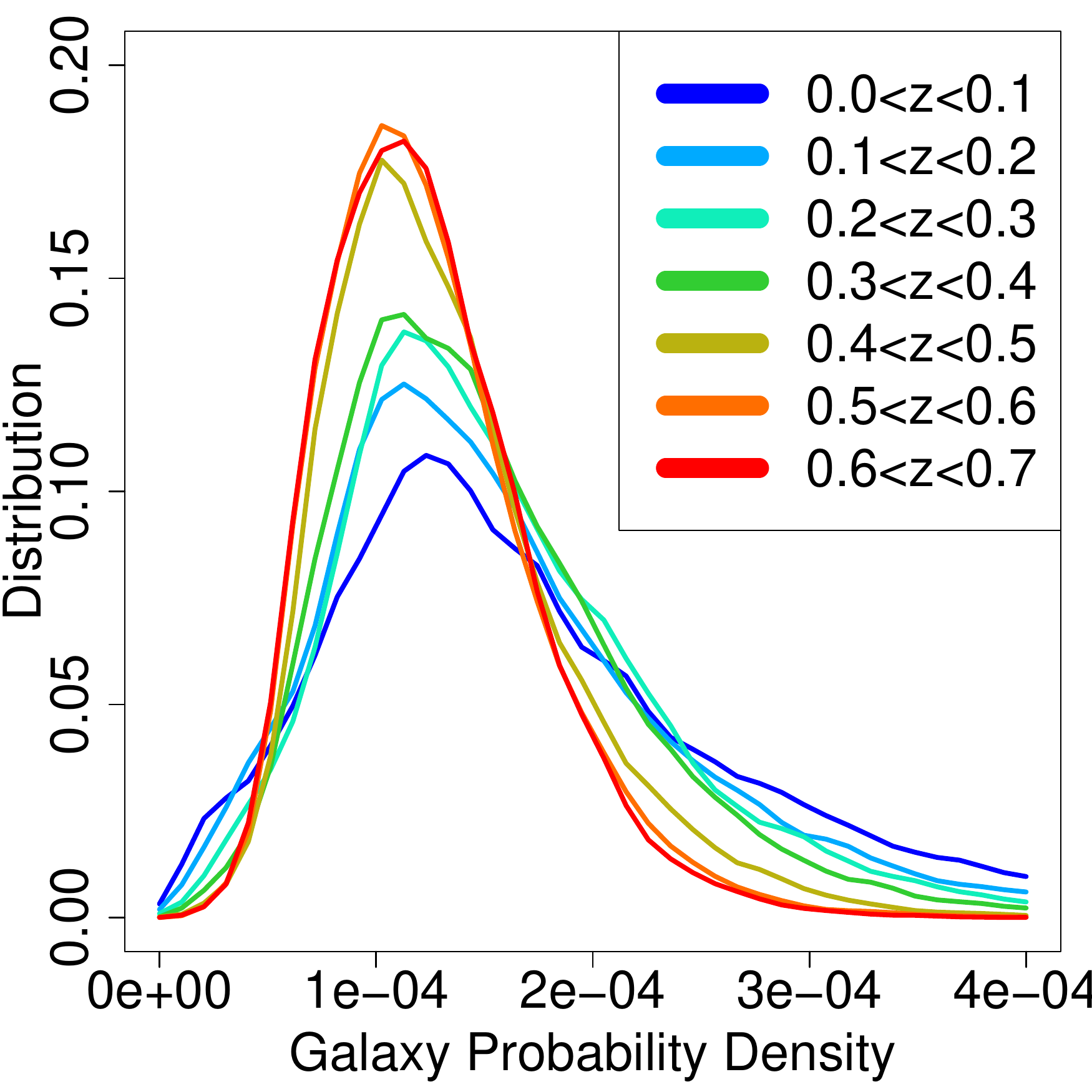}
\caption{Distribution of density profile of filaments at different redshifts.
The distribution of density profile is right-skewness 
at low redshift, indicating that galaxy density on each filament point
is in general higher than those filaments at high redshift.
}
\label{fig::F_density}
\end{figure}

\begin{table*}
\center
\begin{tabular}{l l l}
\hline
Notation & Definition & Comment\\
\hline
$\ra$& Right Ascension   & \\
$\dec$& Declination   &\\
$\zlow$& Redshift  & $\zlow\leq z< \zlow+0.005$\\
$\density$&Galaxy density at each filament point  &\\
$\HH$ & High density indicator & $1$: located at high density regions\\
$\UM$ & Uncertainty (Error)  & \\
$\vra$ & Direction of filament  &\\
$\vdec$ & Direction of filament  &\\
\hline
\end{tabular}
\caption{Definition of variates in the filament map file.}
\label{tab::map}
\end{table*}

Our filament map catalogue\footnote{The catalogue can be downloaded from 
\url{https://sites.google.com/site/yenchicr/catalogue}.}
contains a collection of points on filaments.
These points are obtained via SCMS 
with a uniform grid as the initial points.
Thus, one can view the points in the
filament maps as a uniformly random sample
on all filaments.
Each filament point has seven variables
as listed in Table \ref{tab::map}.
The first two $( \ra,\dec)$ are the location 
within that slice, and $\zlow$ 
is the indicator (as well as the lower bound of redshift) 
for the slice.

The $\density$ is the galaxy probability density
within each slice under the $(\ra,\dec)$-coordinate at each filament point
(the KDE is used to estimate the density).
Thus, the total probability within each slice sums to 1.
The \emph{density profile}  $\Phi_F\equiv \Phi_F(z)$
is the distribution of $\density$ value at each filament point within the same slice
and it
evolves with redshift.
We compare the density profile within different redshift regions in 
Figure~\ref{fig::F_density}.
An advantage for using density under $(\ra,\dec)$-coordinates
is that we do not have to renormalize the probability density because
the size of each slice remains approximately the same.
If we use ordinate cartesian coordinates, the size of each slice
increases when the redshift increases.
As can be easily seen,
at higher redshift, galaxy densities at filament points 
tends to be lower.
The quantity $\HH$ is a high-density indicator
and is related to the RMS density.
If the density of a given filament point is above the RMS density, $\HH=1$, otherwise
it is $0$.

The quantity $\UM$ is the $1\sigma$ uncertainty (error) for detected filaments.
We measure the error for filaments
by bootstrapping the SDSS data 100 times.
Further details may be found in \cite{2014arXiv1406.5663C}.

The last two table's entries are the orientation of
filaments at each filament point.
We use the density gradient at each point on filaments
as a proxy to the direction. 
This proxy is known to be stable \citep{Eberly1996}.

\begin{figure*}
\centering
	\includegraphics[width=5 in]{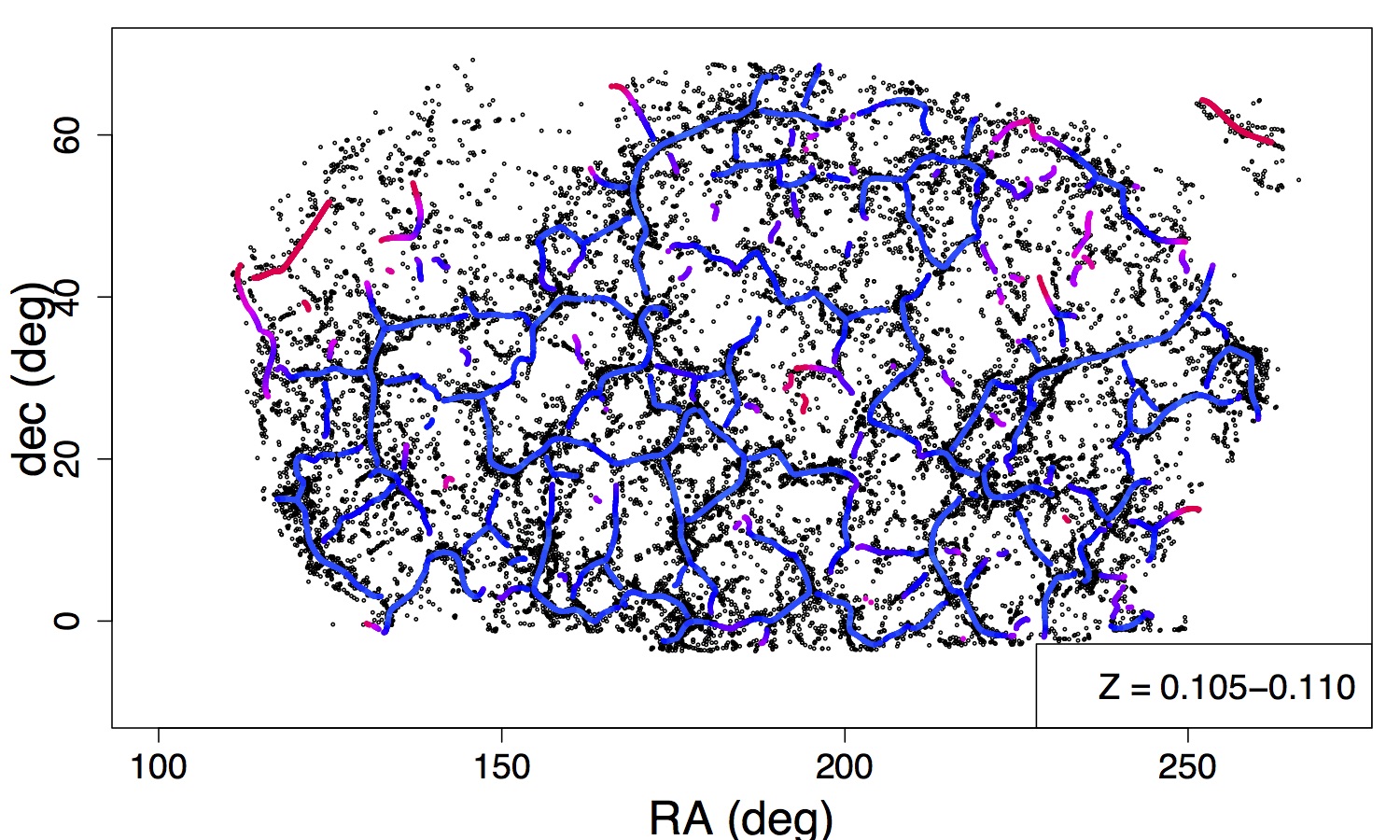}
	\includegraphics[width=5 in]{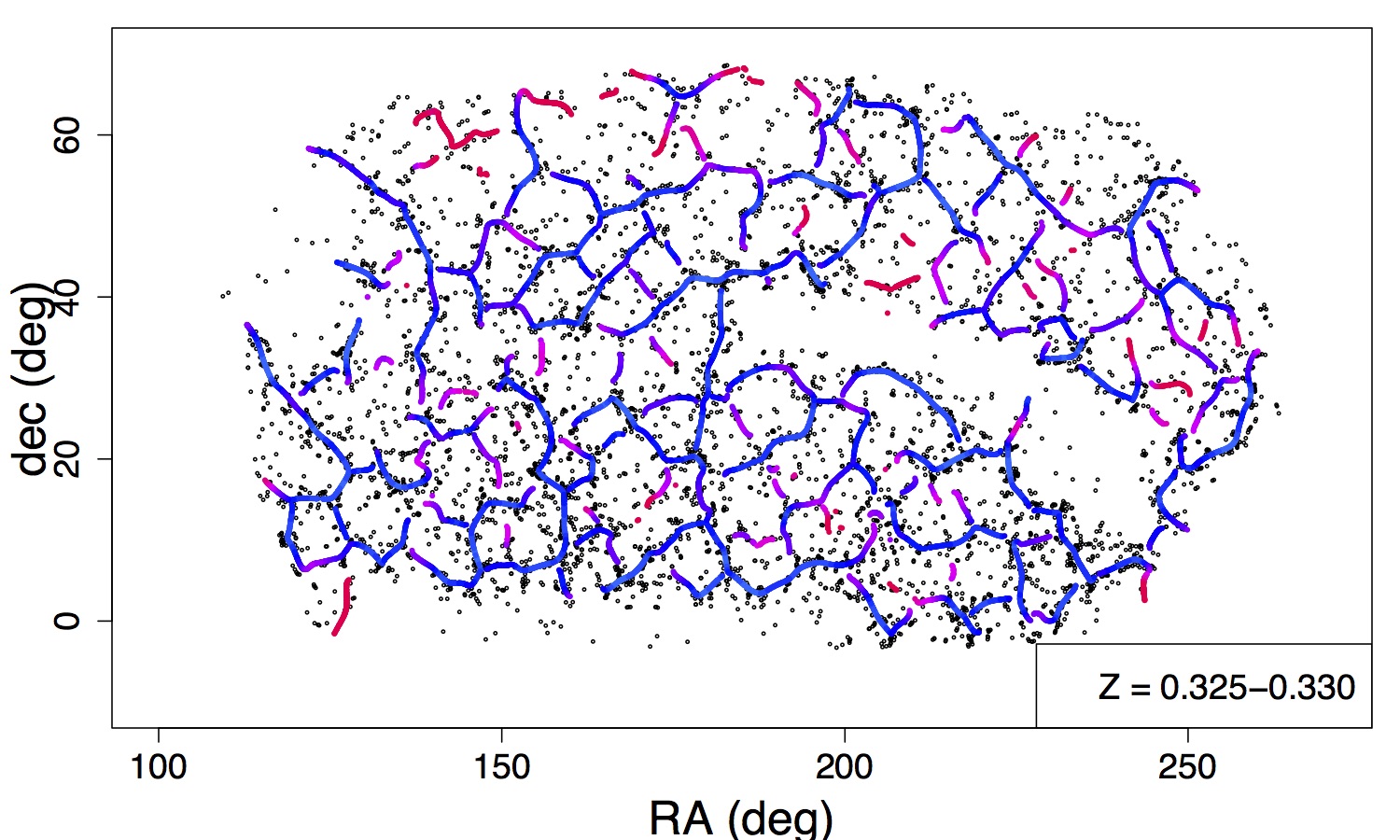}
	\includegraphics[width=5 in]{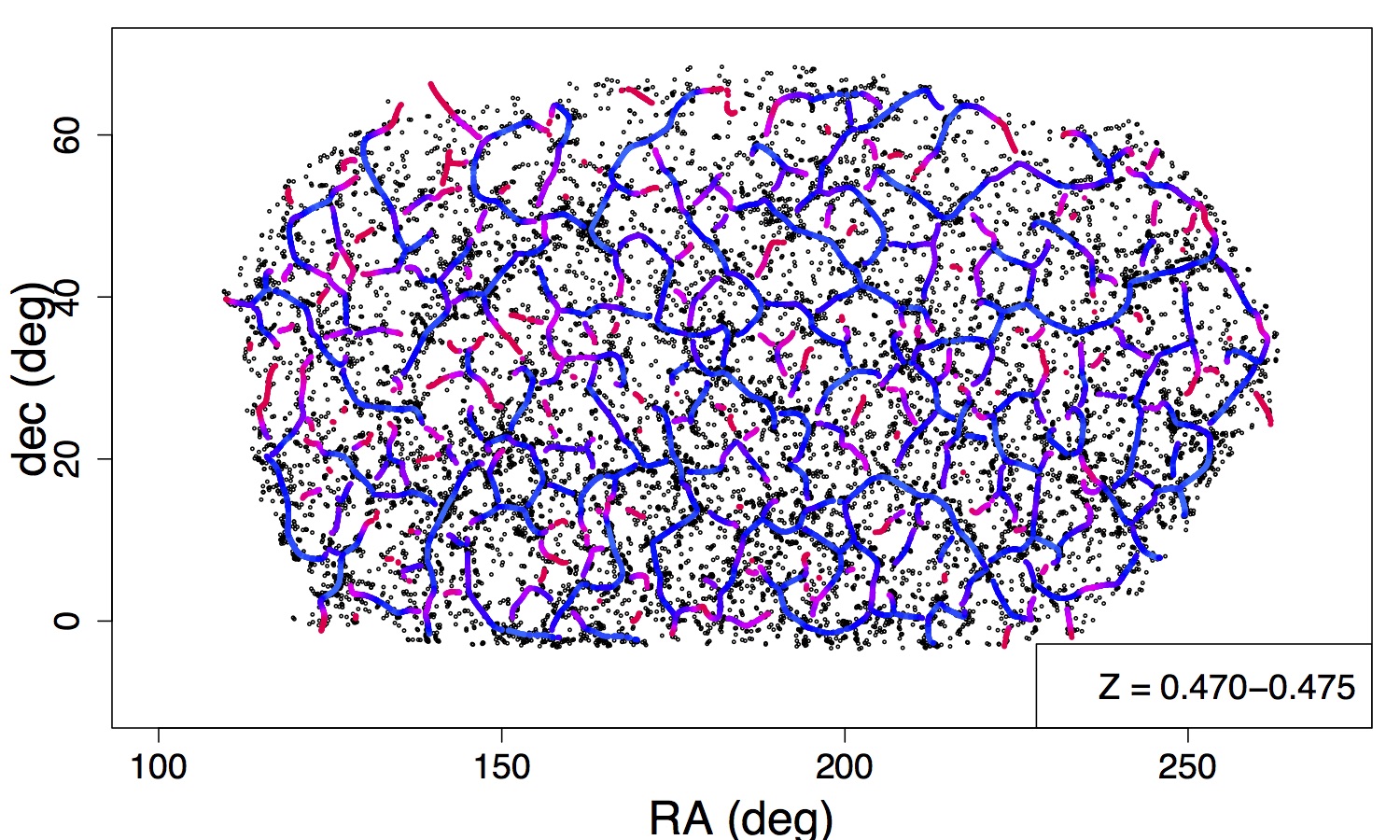}
\caption{Examples of the uncertainty measures of the filament maps.
From top to bottom: $z=0.105-0.110$ (NYU MGS), 
$0.325-0.330$ (LOWZ) and $0.470-0.475$ (CMASS). 
We use color to visualize the amount of uncertainty for filament detection (red $=$ high uncertainty).
Note that the color is relative uncertainty within each slice.}
\label{fig::UMonFM}
\end{figure*}

\section{Filaments at Different Redshifts}	\label{sec::redshift}
The filament maps at each redshift are used to
construct a summary file\footnote{See \url{https://sites.google.com/site/yenchicr/catalogue}. }
that contains information about filaments
at different redshifts.
This file consists of a $130\times 17$ array.
Each row corresponds to a particular slice of the Universe
and each column provides information about that slice.
We describe all 17 variables in the file in Table \ref{tab::info}.

\begin{figure*}
\centering
	\includegraphics[height=2 in, width=2 in]{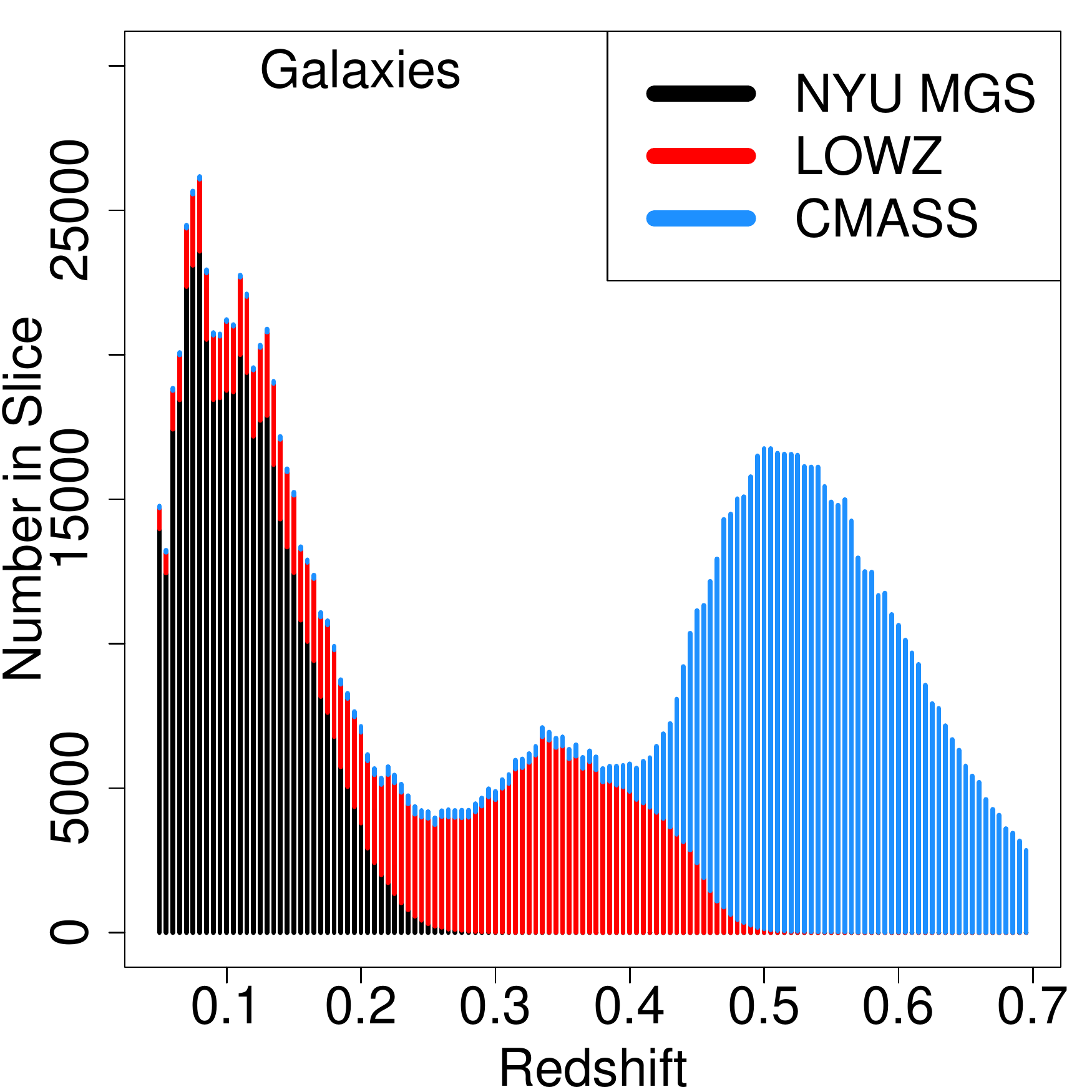}
	\includegraphics[height=2 in, width=2 in]{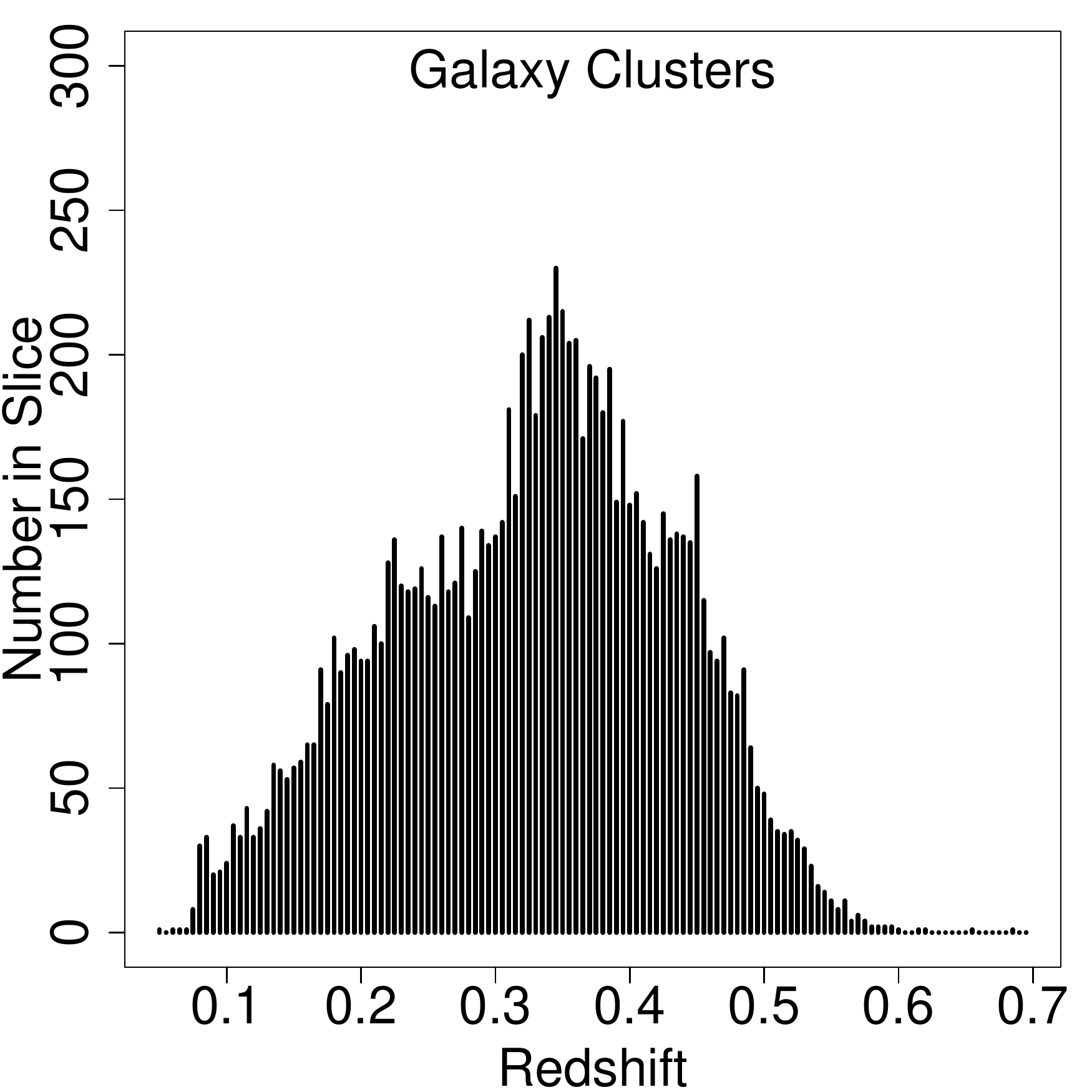}
\caption{{\bf Left:} 
Number of galaxies within each slice.
At boundaries of two catalogues, the number of galaxies
per slice is small.
{\bf Right:} Number of galaxy clusters from reMaPPer catalogue within each slice.
The majority of reMaPPer clusters is in the regions of LOWZ sample.
}
\label{fig::G_number}
\end{figure*}

\begin{figure*}
\centering
	\includegraphics[height=2 in, width=2 in]{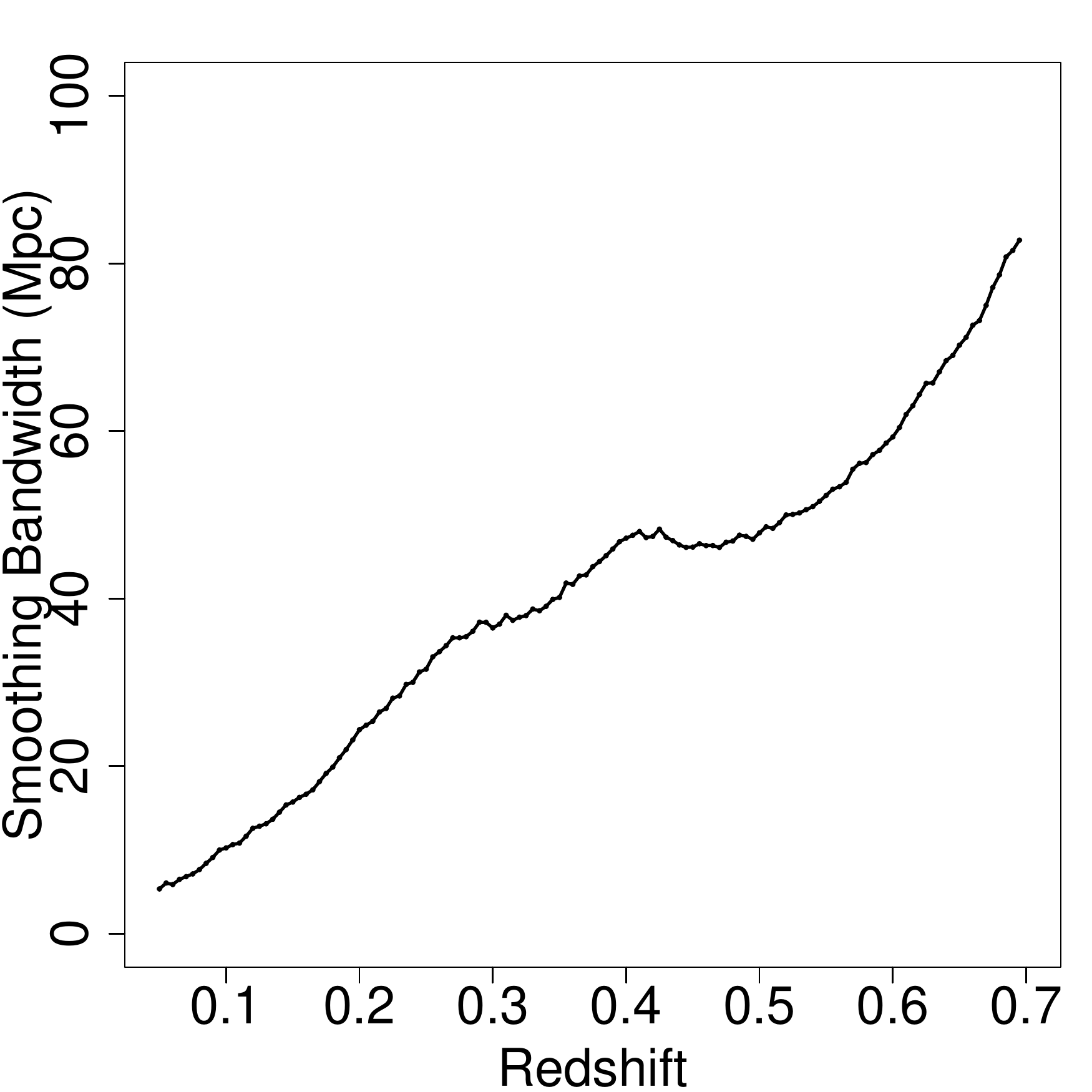}
	\includegraphics[height=2 in, width=2 in]{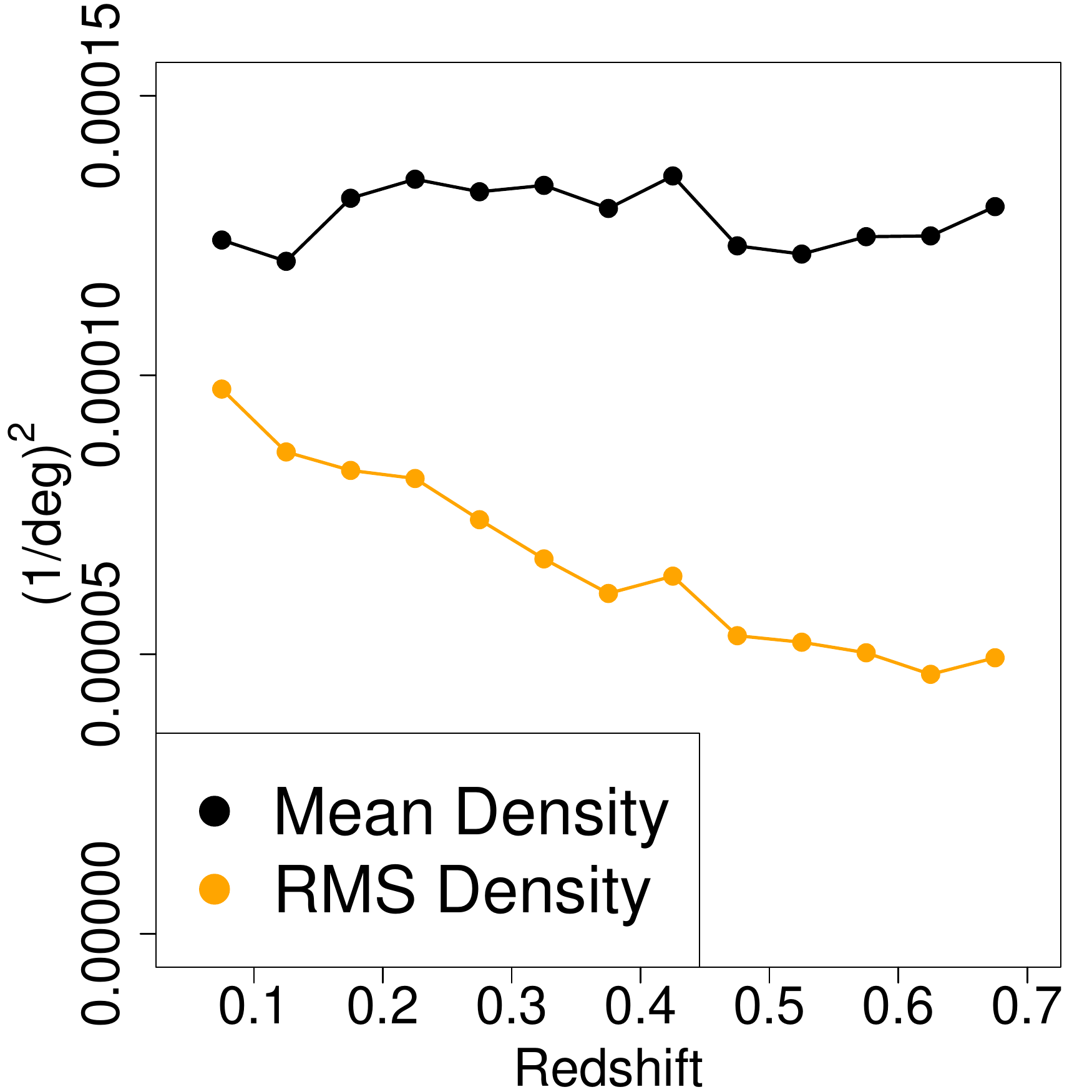}
	\includegraphics[height=2 in, width=2 in]{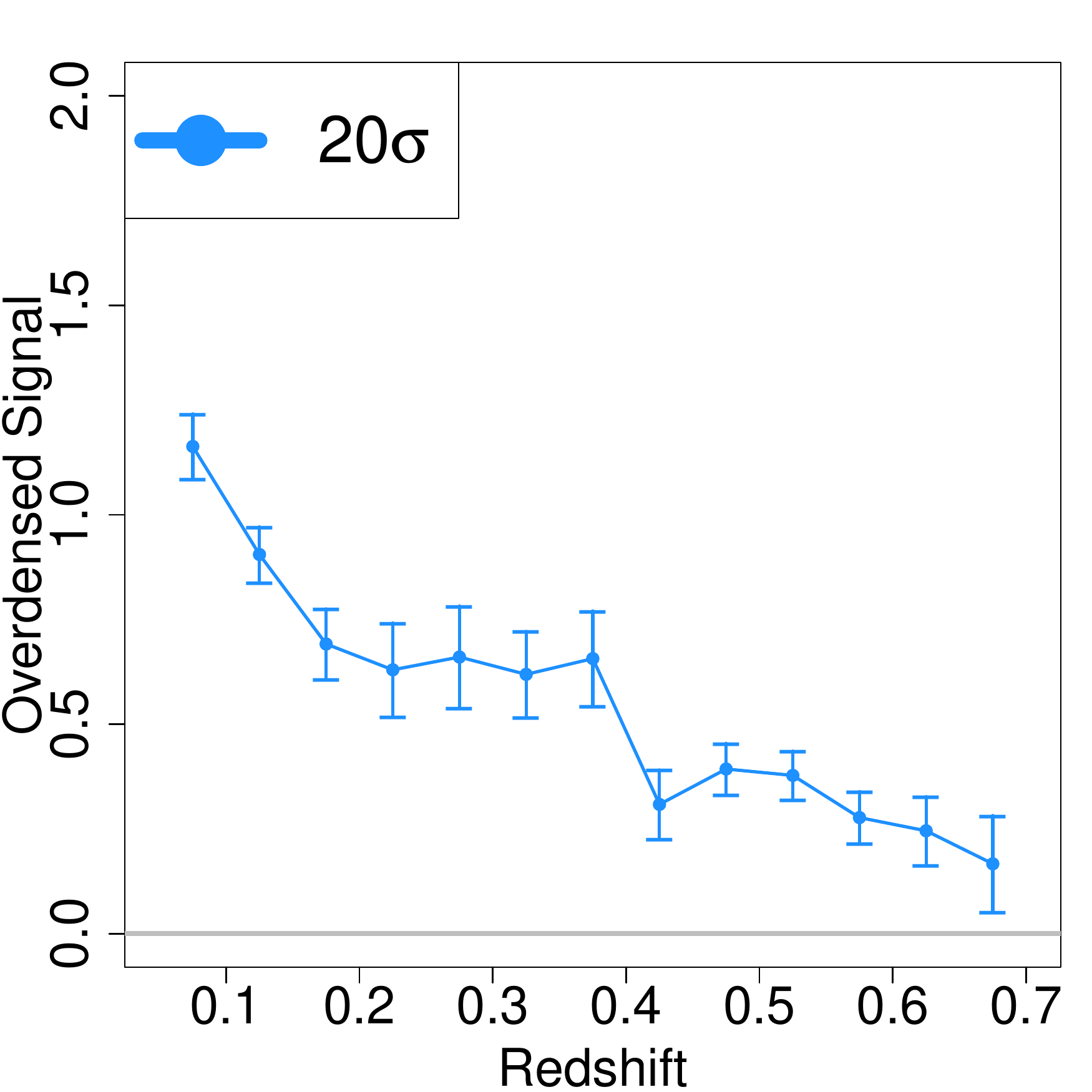}
\caption{{\bf Left:} 
Smoothing bandwidth over the redshift range $0.05-0.70$.
We must apply a larger smoothing bandwidth 
for data in higher redshift since the number density decreases.
{\bf Center:} The mean and the RMS (probability) density 
as a function of redshift.
In generally, the mean density does not vary too much as the redshift
changes. The RMS density, however, decreases when the redshift increases.
See {\S}\ref{sec::redshift} for discussion about possibilities for this pattern.
{\bf Right:} The overdensed signal at different slices. 
The overdensed signal is the average density on all filament points within a slice
minus $\pmean$ and divide $\prms$.
The over-density measures the quality of how filament trac
high density regions.
The decreasing pattern might come from higher errors for detecting filament
at higher redshift or the side effect from smoothing. See {\S}\ref{sec::redshift} for more details.
}
\label{fig::info}
\end{figure*}

\begin{figure*}
\centering
	\includegraphics[height=2 in, width=2 in]{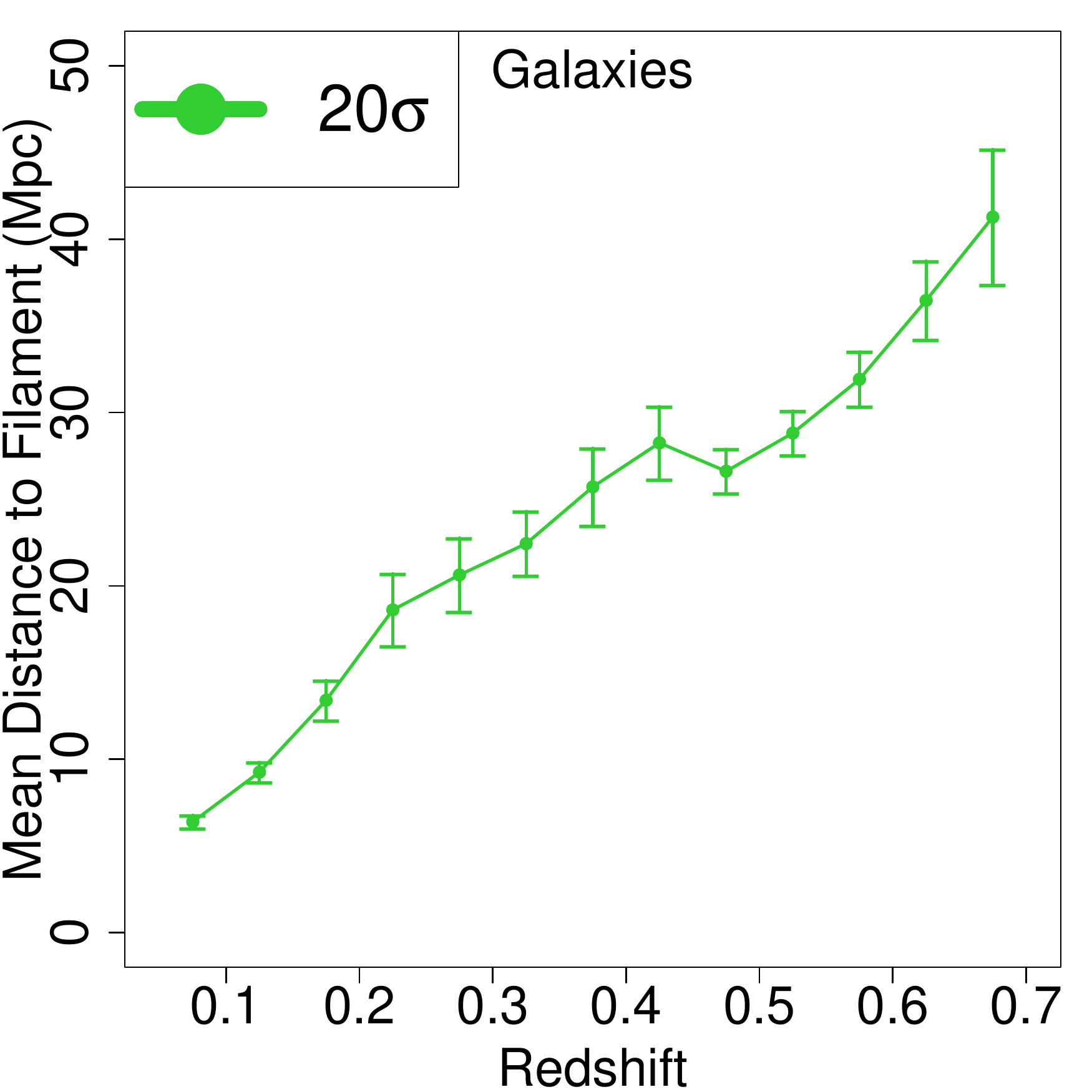}
	\includegraphics[height=2 in, width=2 in]{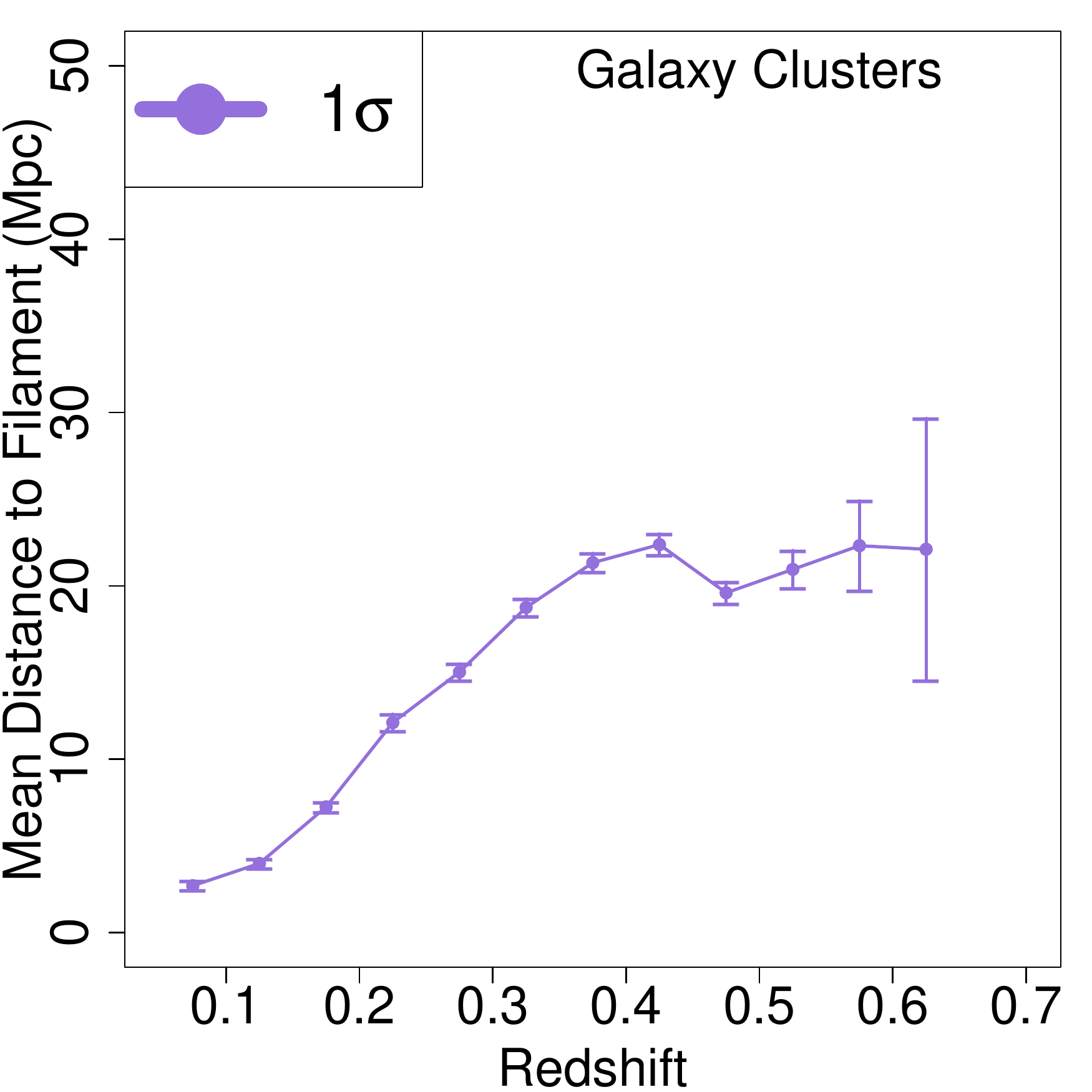}
	\includegraphics[height=2 in, width=2 in]{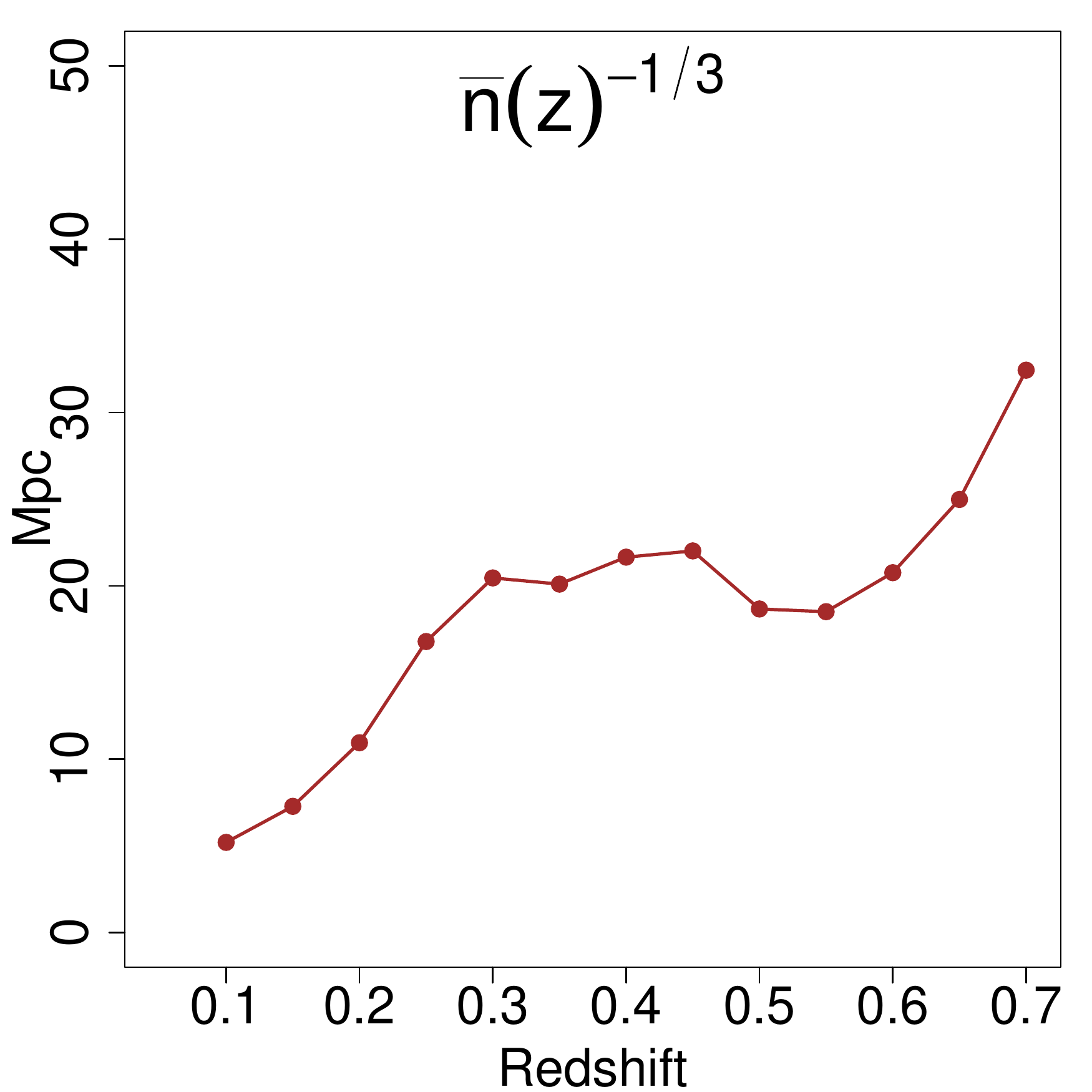}
\caption{
{\bf Left :} 
The distance to filaments from galaxies.
The displayed errors for left panel
are multiplied by a factor of 20 to 
show the minuscule error.
{\bf Center:}
The distance to filaments from galaxy clusters.
Comparing the left panel to the center panel, we see that clearly
clusters are closer to filaments than a randomly select galaxy.
{\bf Right:} The inverse of the cube root of number density $\bar{n}(z)$.
This quantity has the unit of distance and is generally proportional 
to the average distance between galaxies.
Distances to filaments from both galaxies and clusters
have a similar trend as $\bar{n}(z)^{-1/3}$.
This reveals that the increasing pattern in redshift is due to the change in
number density.
}
\label{fig::info2}
\end{figure*}


\begin{table*}
\center
\begin{tabular}{l l l l}
\hline
Notation & Definition  &Unit& Remark\\
\hline
$\zlow$& Redshift value t && $\zlow\leq z< \zlow+0.005$\\
$\N$& Galaxy number  &&\\
$\NGC$& Galaxy cluster number  && \\
$\h_{deg}$&Smoothing bandwidth  &degree&\\
$\h_{mpc}$&Smoothing bandwidth  &Mpc&\\
$\pmean$ & Mean galaxy density  &degree$^{-2}$&\\
$\prms$ & RMS of galaxy density  &degree$^{-2}$&\\
$\Fdensity$ & Mean galaxy density on filaments  &degree$^{-2}$&\\
$\dF$ & Mean galaxy distance to filaments  &Mpc& \\
$\dF_H$ & Mean galaxy distance to high density filaments  &Mpc& \\
$\dF_{gc}$ & Mean cluster distance to filaments  &Mpc&$-1$: $\NGC=0$ \\
$\dF_{gc, H}$ & Mean cluster distance to high density filaments  &Mpc& $-1$: $\NGC=0$ \\
$\UM_{Q1}$ & First quantile for uncertainty of filaments &Mpc&\\
$\UM_{med}$ & Median uncertainty of filaments  &Mpc&\\
$\UM_{avg}$ & Mean uncertainty of filaments  &Mpc&\\
$\UM_{Q3}$ & Third quantile for uncertainty of filaments  &Mpc&\\
$\UM_{rms}$ & Uncertainty fluctuation (RMS) of filaments  &Mpc&\\
\hline
\end{tabular}
\caption{Definition of entries in the catalogue-description file.}
\label{tab::info}
\end{table*}

The first variable ($\zlow$)
is the lower limit on redshift of that slice.
Each slice contains the region
$$
\zlow\leq z<\zlow+0.005.
$$
The second variable ($\N$) is the number of galaxies 
within these regions.
$\NGC$ is the number of galaxy clusters from redMaPPer catalogue
\citep{2014ApJ...785..104R,2014ApJ...783...80R}
within the slice.
We use only the clusters with spectroscopic redshifts
that are within the mask of each SDSS catalogue.
Figure~\ref{fig::G_number} shows $\N$ and $\NGC$
at different redshifts.
The left panel displays the galaxy sample size from three samples:
the NYU main galaxy sample (black),
the LOWZ sample (green)
and the CMASS sample (blue).
The right panel presents the number of clusters at each slice.

The number $\h$ is the smoothing bandwidth used
in density reconstruction and filament detection \cite{2015arXiv150105303C}.
The left panel of Figure~\ref{fig::info} shows the
smoothing bandwidth at different redshifts.
We select $\h$ according to the reference rule in appendix of \cite{2015arXiv150105303C},
which depends on the RMS of the density.
$\h$ increases as the redshift increases
because, at high redshift, the number density of galaxies is
small so we need to enforce a strong degree of smoothing
to detect filaments.
The trend of $\h$ is similar to the inverse of the cube root of number density;
see the right panel of Figure~\ref{fig::info2}.

The two variables $\pmean,\prms$
are the mean overdensity and the RMS density.
The RMS measures the density fluctuation of $p(x)$
and is used in the thresholding procedure of the SCMS algorithm \citep{2015arXiv150105303C}.
We write $\pmean(z)\equiv\pmean$ and $\prms(z)\equiv\prms$ 
since the mean density and RMS density change as the redshift changes.
The center panel of Figure~\ref{fig::info}
displays $\pmean(z)$ and $\prms(z)$.
It is clear that $\prms(z)$ decreases as redshift increases
while the mean density $\pmean(z)$ remains roughly the same.
These effects occur for two reasons. First, density fluctuations
are smaller at early times (higher redshifts); second,
the smoothing parameter $\h$ is larger at higher redshifts
so that the density estimate is strongly smoothed, reducing the 
amplitude of fluctuations.

The quantity $\Fdensity$ is the 
average density profile of filaments at the given slice
and is related to the result in Figure~\ref{fig::F_density}, which
shows the distribution of density profile
at wide-redshift regions. 
The difference between $\Fdensity$ and $\pmean$
is that $\Fdensity$ is the average density value on filaments only
while $\pmean$ is the average density value on the whole region of observation.
The right panel of Figure~\ref{fig::info2}
presents the over-density for filaments, which
is defined as
$$
S(z) = \frac{\Fdensity(z)-\pmean(z)}{\prms(z)}.
$$
The over-density shows how the filaments trace high density regions.
If $S(z)$ is large, then most filaments within this slice trace
high density regions.
As can be seen, the over-density for filaments
decreases as redshift increases,
implying that filaments do not trace high density regions
so well at the high redshift range.
There are many possible explanations for this result.
At higher redshift regions, the number density
is lower so that our filament reconstruction has larger errors.
Another possibility is that, at higher redshift, the smoothing parameter $\h$
is also larger, which
flattens the density fluctuation.


The quantity $\dF$ is the average distance from all galaxies
to filaments within the specified slice.
The related quantity $\dF_{H}$ is analogous to $\dF$,
but uses the distance to `high-density' filaments,
i.e.,
the distance to filament points whose density is above the RMS density.
The Left panel of Figure \ref{fig::info2} displays $\dF$ at each redshift.
The average distance to filaments increases as redshift increases.
This increasing pattern is caused by the change in number density--
the higher redshift regions  generally have lower number density.
To demonstrate how number density affects the average distance,
we provide the inverse of the cube root of number density at each redshift
at right panel of Figure \ref{fig::info2};
the pattern in the left and right panels are clearly similar.

The quantity $\dF_{gc}$ is the mean distance to filaments
from galaxy clusters (redMaPPer clusters; \citealt{2014ApJ...785..104R, 2014ApJ...783...80R});
$\dF_{gc, H}$ is similar to $\dF_H$ but is
evaluated at each galaxy cluster. 
The quantity $\dF_{gc,H}$ is the 
mean distance to the high-density filament from galaxy clusters.
If $\NGC=0$, both $\dF_{gc}$ and $\dF_{gc,H}$ are set to be $-1$.
The center panel of Figure~\ref{fig::info2} shows the mean distance of galaxy clusters 
$\dF_{gc}$ under various redshifts.
Basically, $\dF_{gc}$ follows a similar trend as $\dF$
but has a lower value,
indicating that, on average,
clusters are closer to filaments than a randomly selected galaxy.


\begin{figure}
\centering
	\includegraphics[height=2.5 in, width=2.5 in]{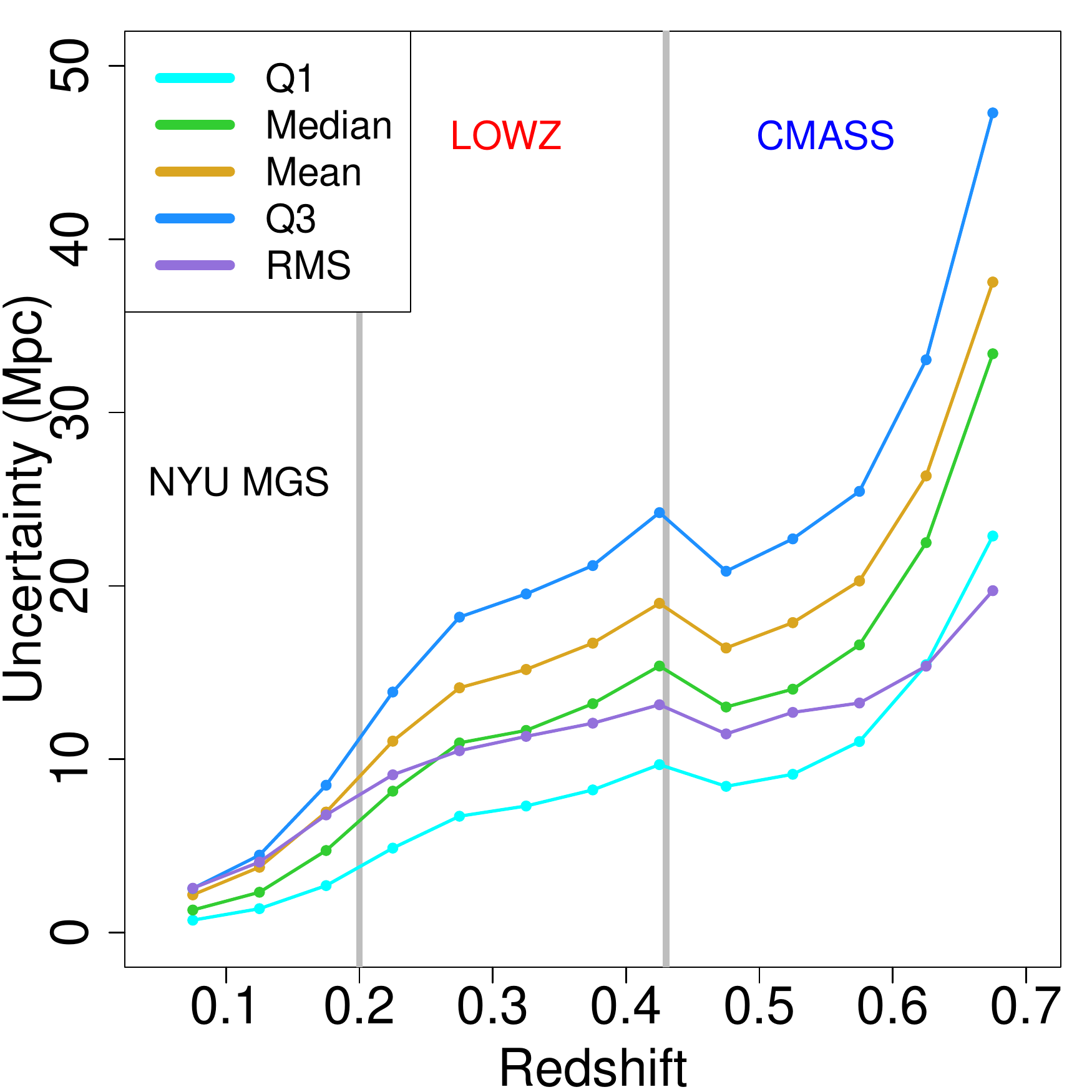}
\caption{Summary statistics showing how filament uncertainty measures
evolve with redshift.
The increasing pattern along redshift for all uncertainty measures
is from the change in number density (c.f. right panel of Figure~\ref{fig::info2}).
}
\label{fig::UM}
\end{figure}

Finally, the five quantities $\UM_{Q1}, \UM_{med}, \UM_{avg}, \UM_{Q3}$ 
and $\UM_{RMS}$ are 
summary statistics for the uncertainty distributes on filaments within each slice.
These quantities are the first quantile (25\%), median, mean, third
quantile (75\%) and root mean square for all the uncertainty values on filament.
The uncertainties are computed using the bootstrap method of
\cite{2015arXiv150105303C}.
The summary at different redshifts
is presented in Figure~\ref{fig::UM}.
The increase of the uncertainty as a function of redshift is due to the change
in number density (c.f. right panel of Figure~\ref{fig::info2}).

\subsection{Filament Evolution} 	\label{sec::FE}

The metric we adopt for quantifying the evolution of filaments
is the ratio of galaxies and clusters within filaments
at different redshifts.
To account for the difference in number density due to the redshift,
we first derive a scaled distance to the nearest filament
for each galaxy (and cluster) using the smoothing parameter and uncertainty measures.
Let $D$ be the distance to filament from a galaxy, and $\pi$
be the nearest point on a filament and
$U$ be the uncertainty measure at $\pi$ 
(the uncertainty measure is defined only for points on filaments).
The scaled distance (to the nearest filaments) from a specified galaxy is defined as
\begin{equation}
S = \sqrt{\frac{D^2+U^2}{h^2}},
\label{eq::S1}
\end{equation}
where $h$ is the smoothing parameter.
We divide the distance by smoothing parameter
so that this scaled distance is comparable from slice to slice
(otherwise for galaxies at lower redshift, $S$ will be much smaller than 
galaxies at higher redshift).
A galaxy (or a cluster) is classified as \emph{within} a filament if
\begin{equation}
S\leq 0.3246.
\label{eq::S2}
\end{equation}
The constant $0.3246$ arises from the density of Gaussian distribution.
Let $\phi(x) = \frac{1}{\sqrt{2\pi}}e^{-x^2/2}$ be the Gaussian distribution.
Then
$$
\frac{\phi(0.3246)}{\phi(0)}\approx 0.9.
$$
If we convolve a true filament with a Gaussian,
the resulting filamentary regions are those points with potential above $90\%$.
i.e., galaxies or clusters within these regions are 
recognized as being `within' filaments.

\begin{figure}
\centering
	\includegraphics[height=2.5 in, width=2.5 in]{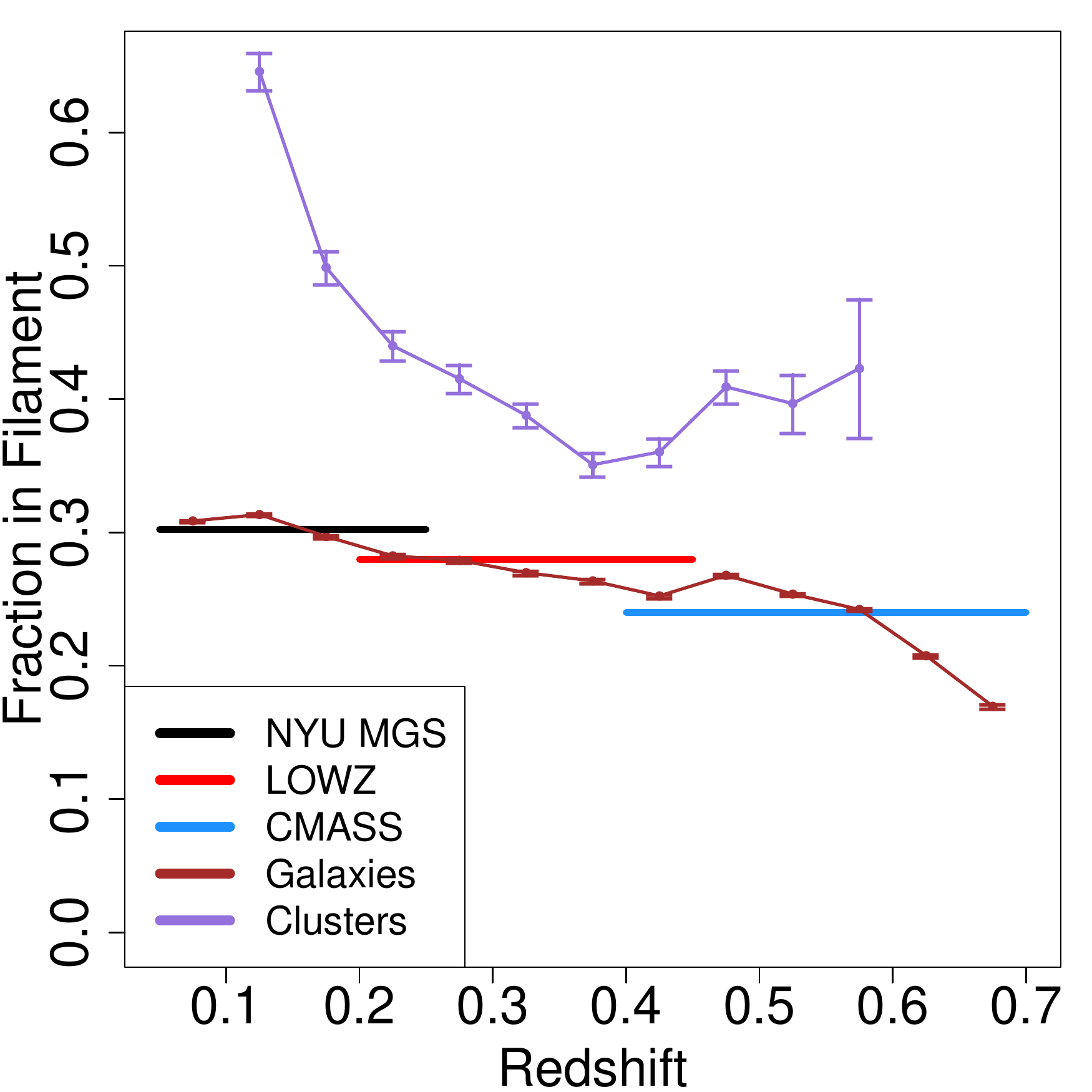}
\caption{Fraction of galaxies and clusters within filaments at different redshifts.
The fraction of clusters within filaments (purple curves)
roughly follows the trend of number density at different redshifts.
In general, our result suggests that roughly $30\%$ galaxies
are in filaments.
}
\label{fig::RatioGC}
\end{figure}

Figure~\ref{fig::RatioGC} displays the proportion 
of galaxies from different catalogues as well as clusters
that are `within' filaments using criterion \eqref{eq::S1} and \eqref{eq::S2}.
The three color bars (black, red and blue)
are the `mean' proportion for NYU MGS, LOWZ and CMASS galaxies.
The brown line is the result for galaxies from all the samples at different redshifts,
and purple lines are the ratios for clusters.
For galaxies, there is a clear decrease with redshifts,
with a small bump at $z\sim0.5$.
This region is the beginning of CMASS sample, so that 
the number density is in fact increasing (see Figure~\ref{fig::G_number}),
therefore
our detection power is increasing. 
The width of error bar at Figure~\ref{fig::UM}
also drops at the $z=0.5$, indicating
the same pattern.
This effect is stronger for galaxy clusters.

Another useful statistic is the proportion of
`stable' filament points.
We classify a filament point as \emph{stable} if
\begin{equation}
\UM \leq \overline{\UM}+ k\UM_{RMS},
\end{equation}
where $\overline{\UM}$ and $\UM_{RMS}$ are the mean
and the root mean square of
the uncertainty over all filament points across every slice.
The number $k$ is the threshold level for defining 
a filament point as stable.

\begin{figure}
\centering
	\includegraphics[height=2.5 in, width=2.5 in]{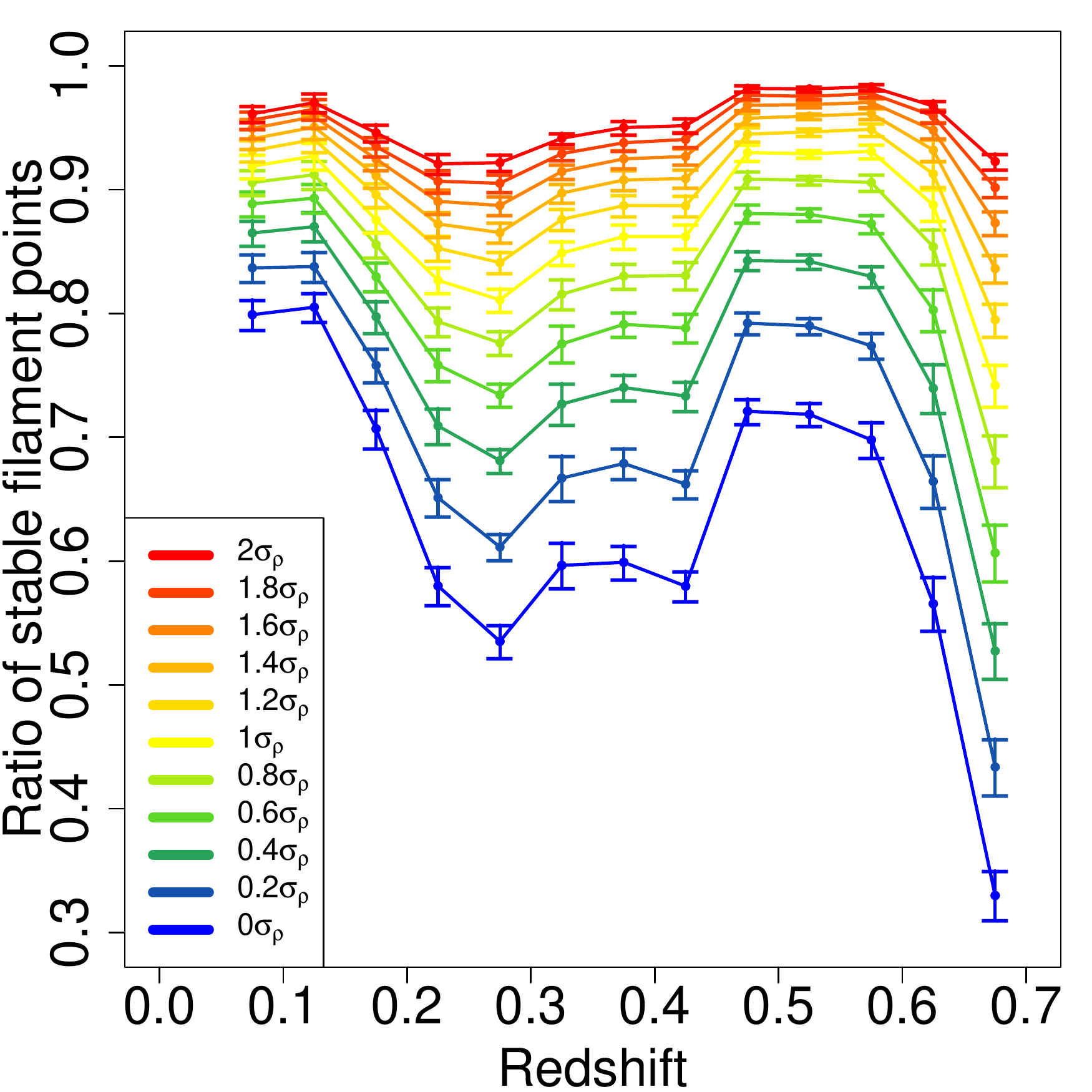}
\caption{Ratio of `stable' filament points at different redshifts under different threshold levels.
Changing the threshold for stability reveals an interesting pattern for the ratio 
of stable filament points, which is in a similar trend of number of galaxies within each slice.
See Section \ref{sec::FE} for further details.
}
\label{fig::RatioSG}
\end{figure}

Figure~\ref{fig::RatioSG} displays
the proportion of stable filaments as a function of redshift
under $k$ ranging from $0$ to $2$.
For all $k$, we see a clear pattern that
the ratio first drops and then increases and drops again.
This phenomenon is even stronger at smaller $k$.
This pattern is similar to that of the number of observations
at each slice (cf. Figure \ref{fig::G_number}).

\subsection{Filament Intersections}
\label{sec::int}

As mentioned in the introduction,
SCMS filaments have an attractive property that
they have good agreement with known galaxy clusters.
\cite{2015arXiv150105303C} demonstrated that most clusters are generally 
close to the detected filaments.
In this section, we identify intersections 
for filaments and compare them to locations of galaxy clusters.

To obtain filament intersections,
we apply a simple algorithm derived from 
metric graph reconstruction \citep{Aanjaneya2012, 2013arXiv1305.1212L},
a method from computational geometry,
to the filaments detected by SCMS
\footnote{We also provide the intersections of filaments in
\url{https://sites.google.com/site/yenchicr/catalogue}.}.
The implementation details can be found in Appendix \ref{sec::int::alg}.

Figure~\ref{fig::FMex} presents an example of applying
this detection algorithm to our filament maps.
The orange color points are intersections.
The detection algorithm clearly successfully
identifies the intersection points,
and most galaxy clusters are close to these points.

\begin{figure*}
\centering
	\includegraphics[height=1.5 in]{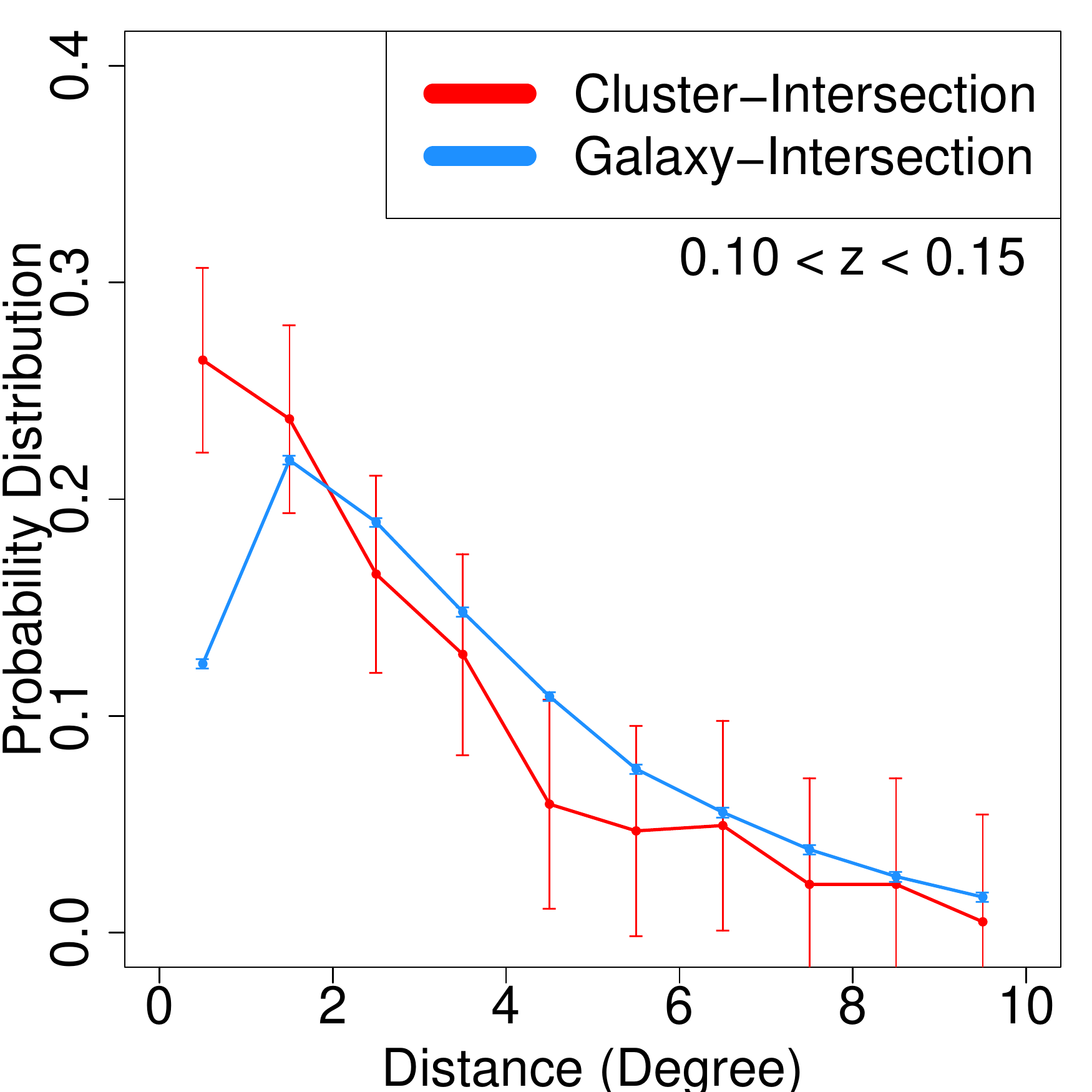}
	\includegraphics[height=1.5 in]{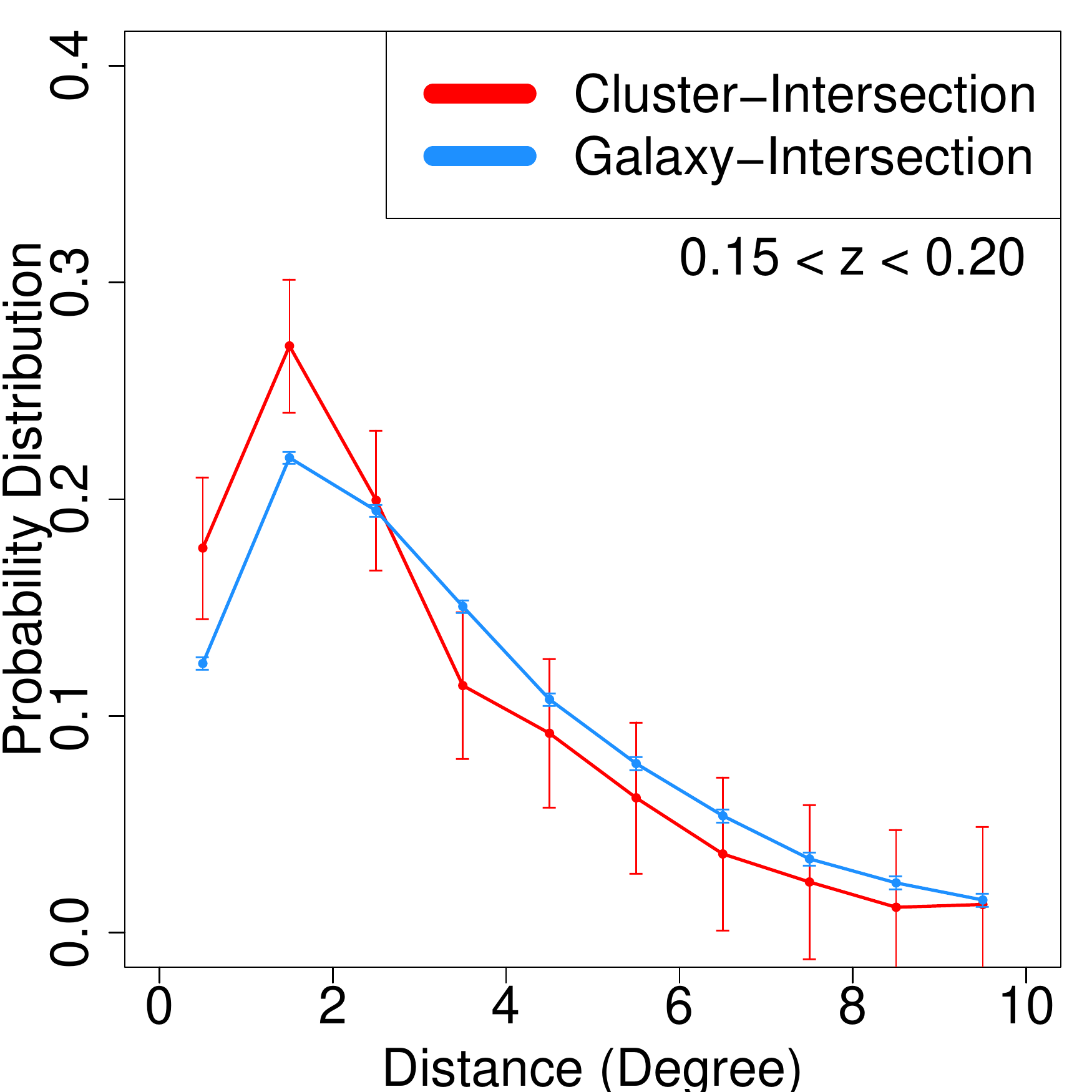}
	\includegraphics[height=1.5 in]{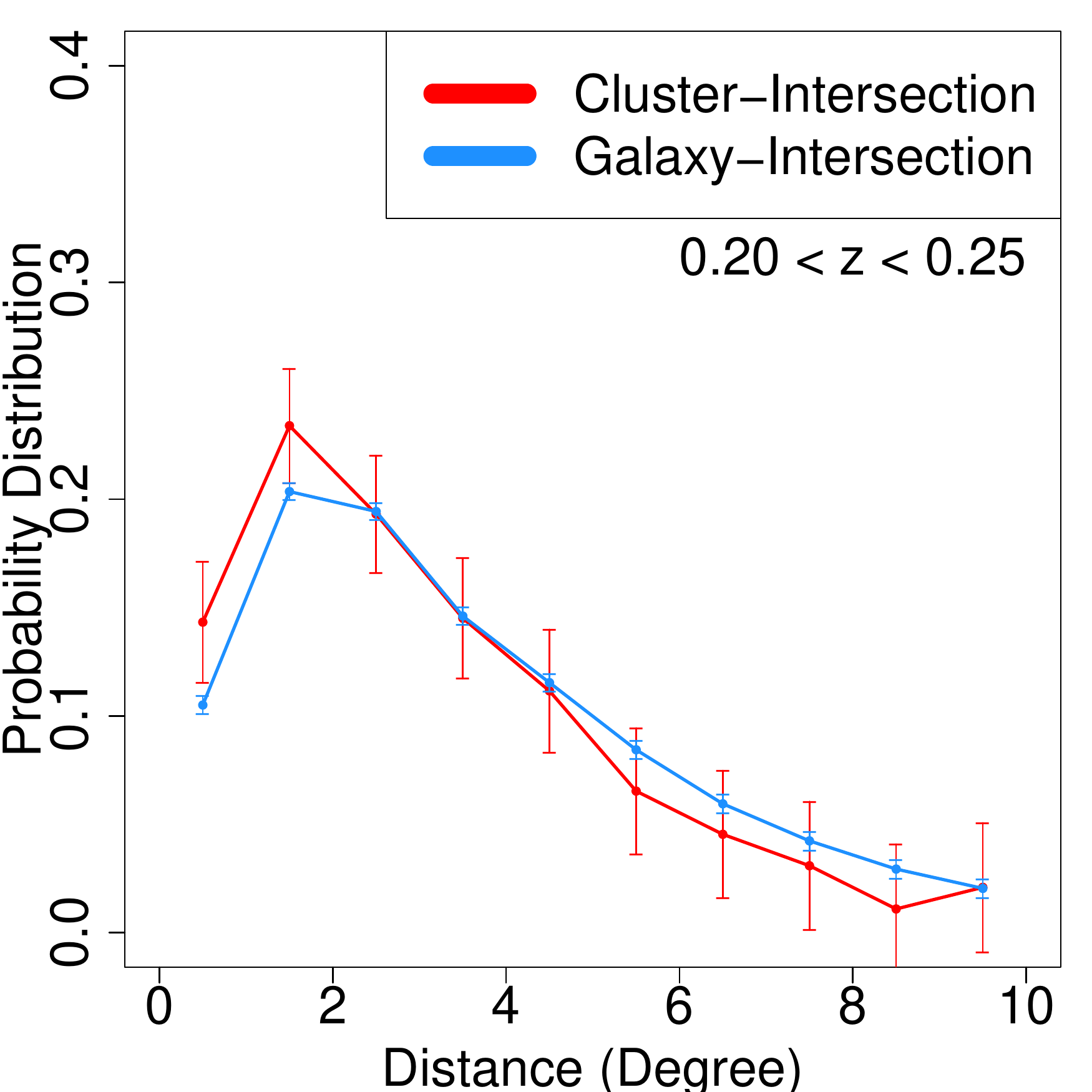}
	\includegraphics[height=1.5 in]{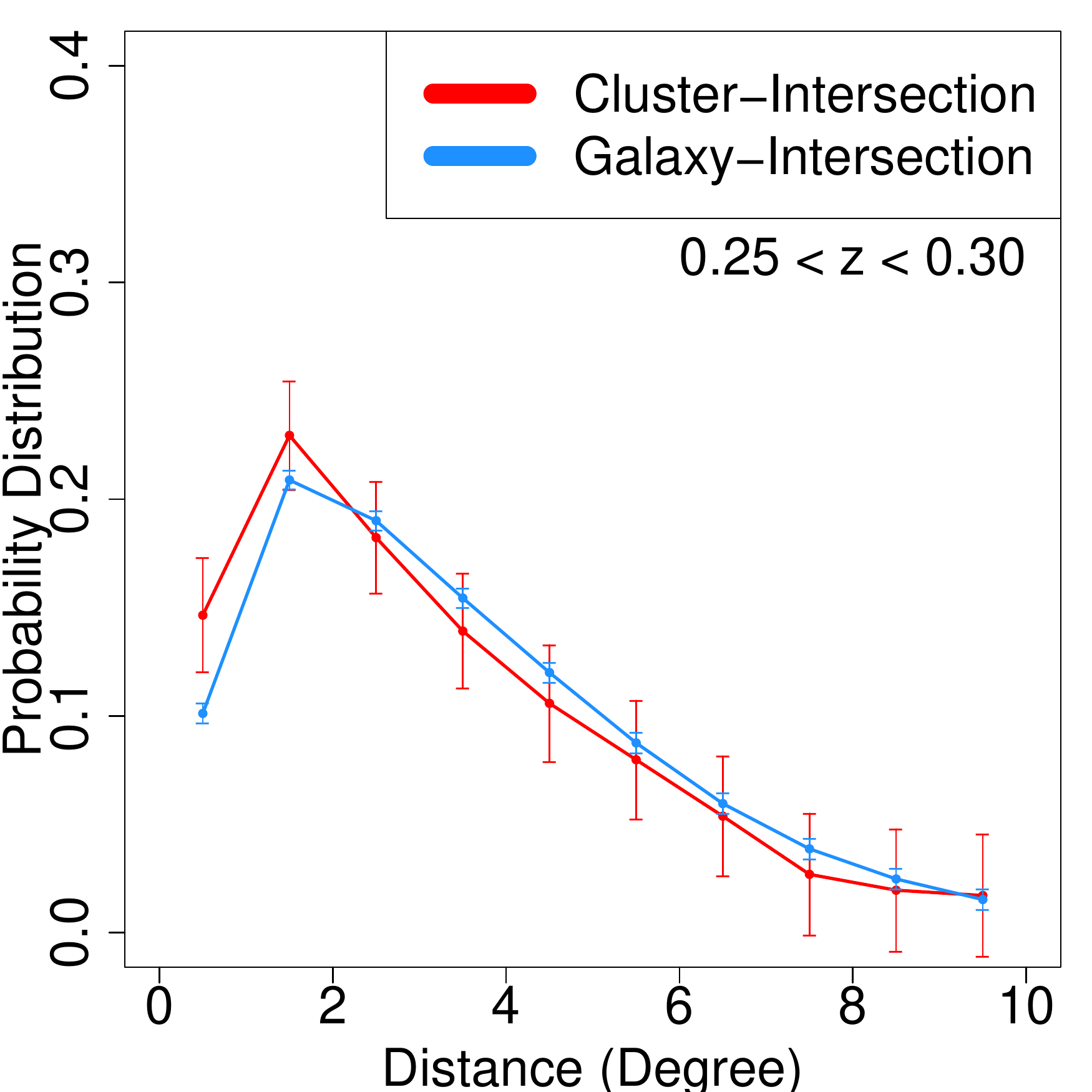}\\
	\includegraphics[height=1.5 in]{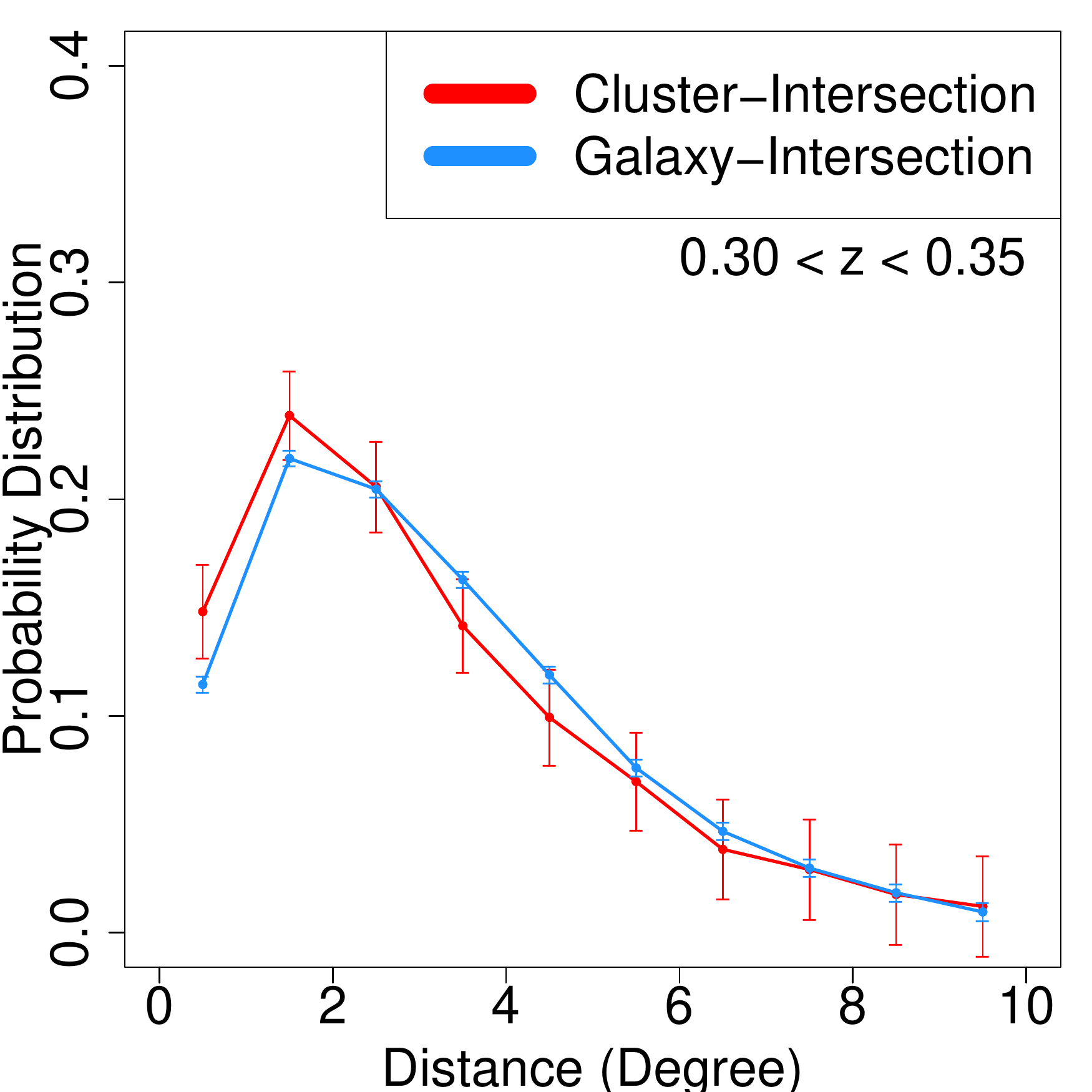}
	\includegraphics[height=1.5 in]{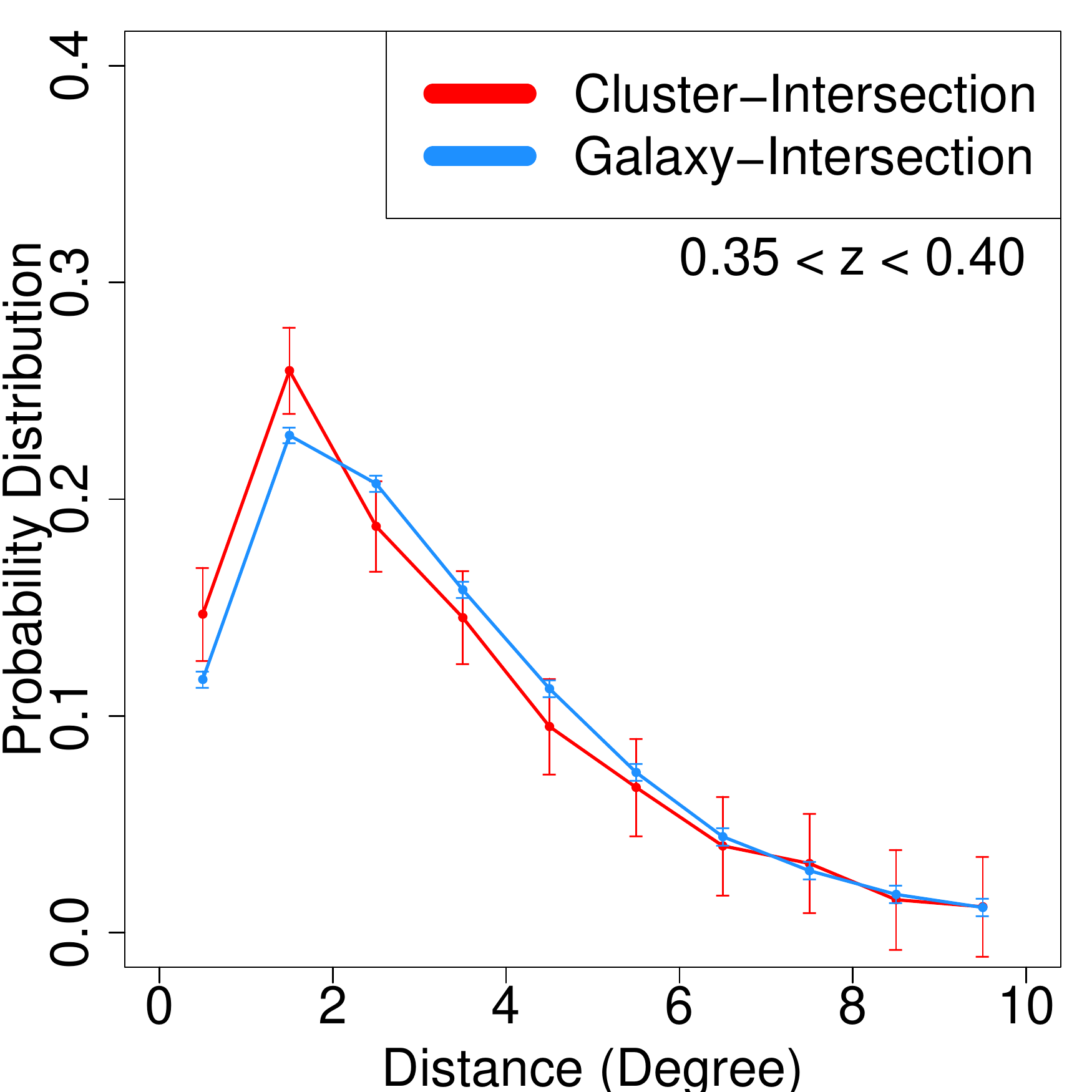}
	\includegraphics[height=1.5 in]{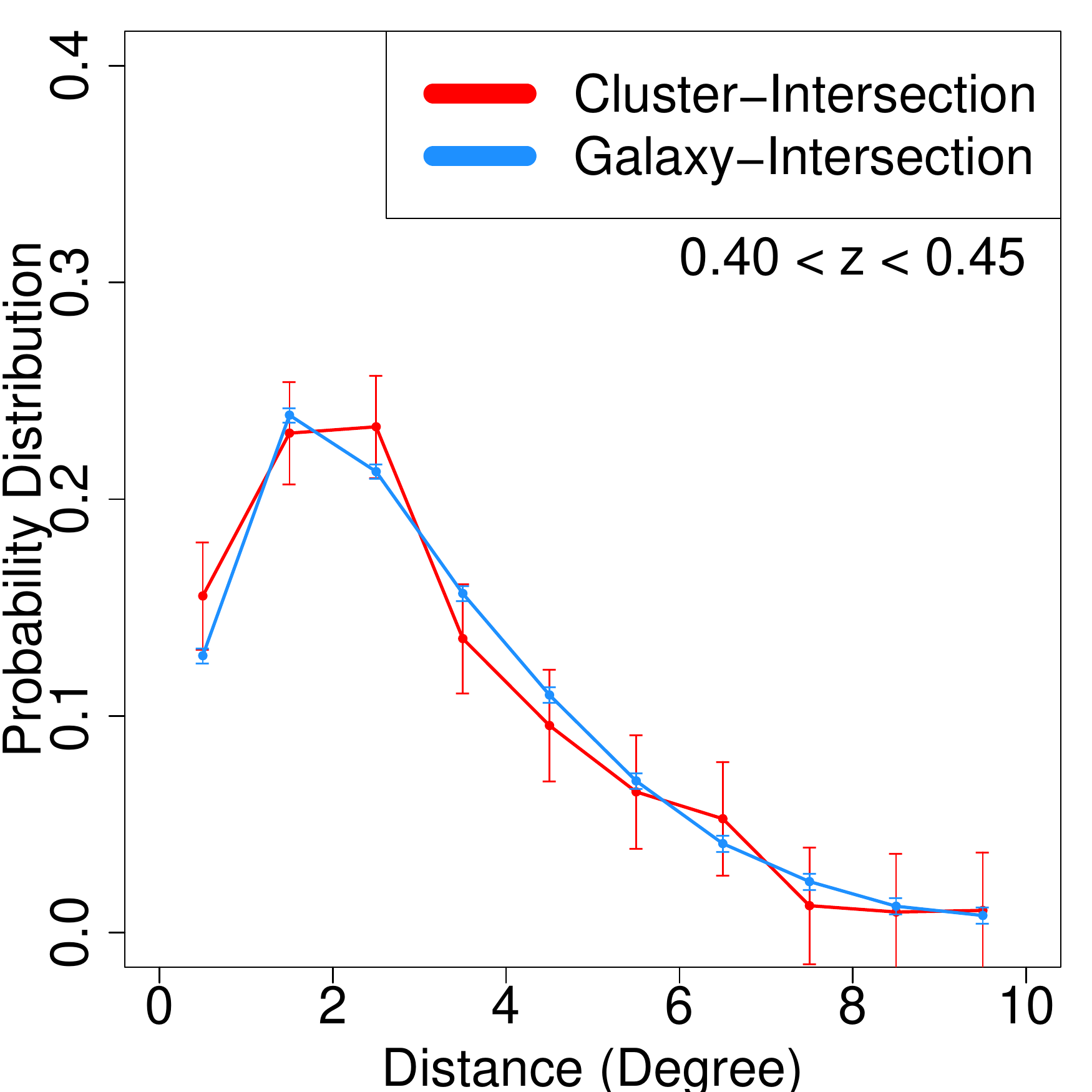}
	\includegraphics[height=1.5 in]{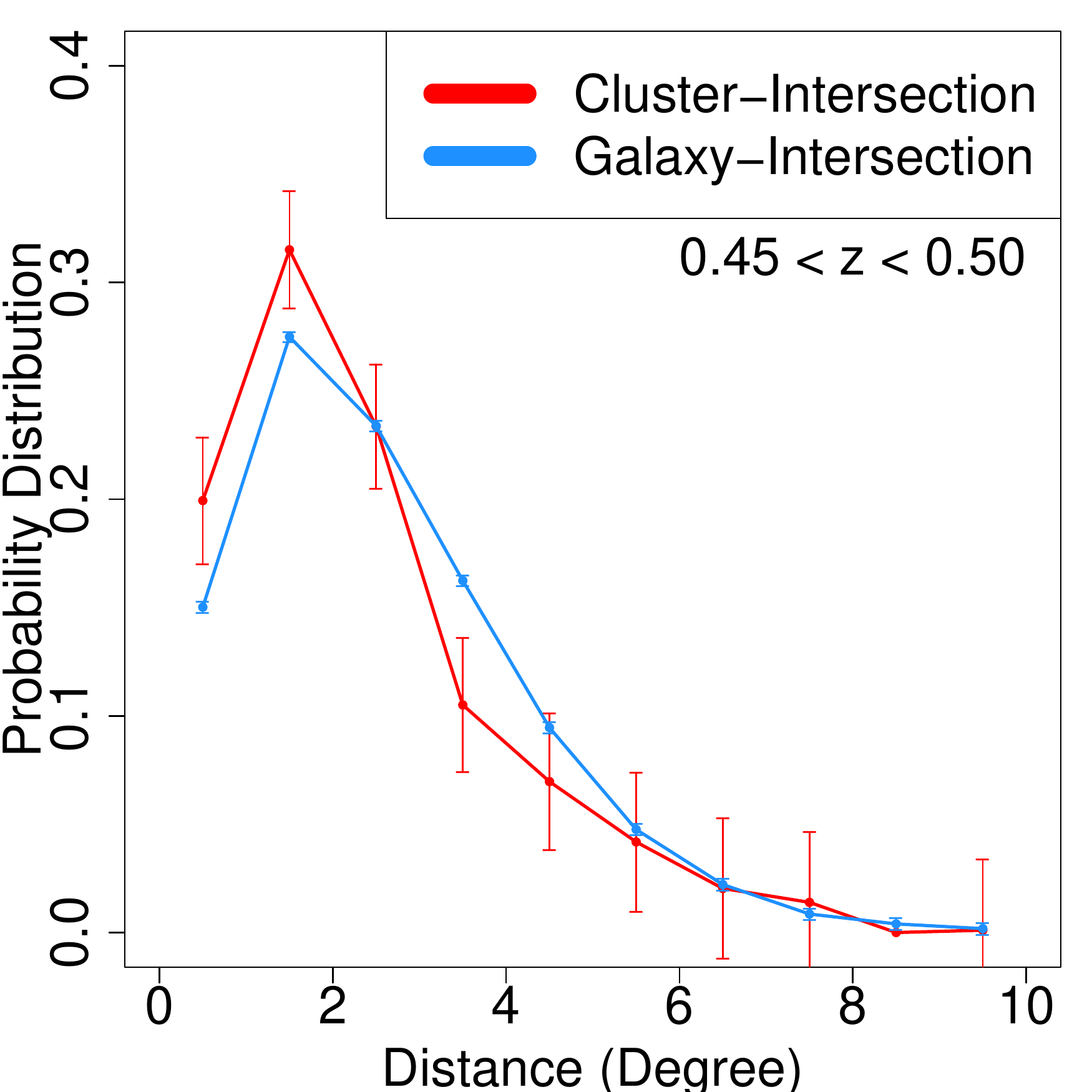}\\
\caption{Distributions of distances from clusters (red) and galaxies (blue) to intersection points.
Each panel is the result for a particular redshift region.
In every panel, we observe that
clusters generally have shorter distances to intersection points
than randomly selected galaxies. 
To make this statement quantitatively, we perform KS test for each pair of distributions.
The result is given in Table \ref{tab::int}.
}
\label{fig::int2}
\end{figure*}

To quantify the closeness of clusters to intersection points,
we compute the distance from galaxy clusters to
the intersection points at different redshifts
and compare this distance statistic
to the distance from a random galaxy point to the intersection.
Figure \ref{fig::int2} shows the distribution of distances
from clusters (red) versus distance from galaxies.
We use the one-sided KS-test to compare the difference in 
distribution; the result is given in Table \ref{tab::int}.
The clusters are significantly closer
to intersections for filaments compared to galaxies.
The worst case (largest p-value) is at $0.4<z<0.45$.
This region corresponds to the boundary between LOWZ and CMASS
samples and is the region with the smallest number density of galaxies 
(cf. Figure \ref{fig::G_number}).
Thus, our filament detection algorithm lacks statistical power at this region,
so it is expected that the p-value is largest here.

\begin{table}
    \begin{tabular}{|l|l|l|l}
    \hline
    Redshift    & $p$-value & Redshift    & $p$-value \\ \hline
    0.100-0.150 & $2.79\times 10^{-11}$       & 0.300-0.350 & $2.68\times 10^{-6}$       \\ \hline
    0.150-0.200 & $2.55\times 10^{-9}$       & 0.350-0.400 & $2.16\times 10^{-7}$       \\ \hline
    0.200-0.250 & $2.58\times 10^{-8}$       & 0.400-0.450 & $1.28\times 10^{-2}$       \\ \hline
    0.250-0.300 & $4.47\times 10^{-8}$       & 0.450-0.500 & $4.95\times 10^{-8}$       \\ \hline
    \end{tabular}
\caption{Significances generated from a one-sided, two-sample KS test, for
the null hypothesis that galaxy clusters lie at the same average distance from
intersections as field galaxies. 
$p$-value is a statistical quantity to measure the significance.
Typically, the usual rejection rule requires $p<0.05$.
}
\label{tab::int}
\end{table}

\section{Magnitude and Distance to Filaments}
We use the filament maps to investigate the relation between a galaxy's
magnitude and its distance to filaments.
Specifically, we wish to determine whether
galaxies near filaments tend to be more luminous.

\begin{figure*}
\centering
	\includegraphics[height=2 in, width=2 in]{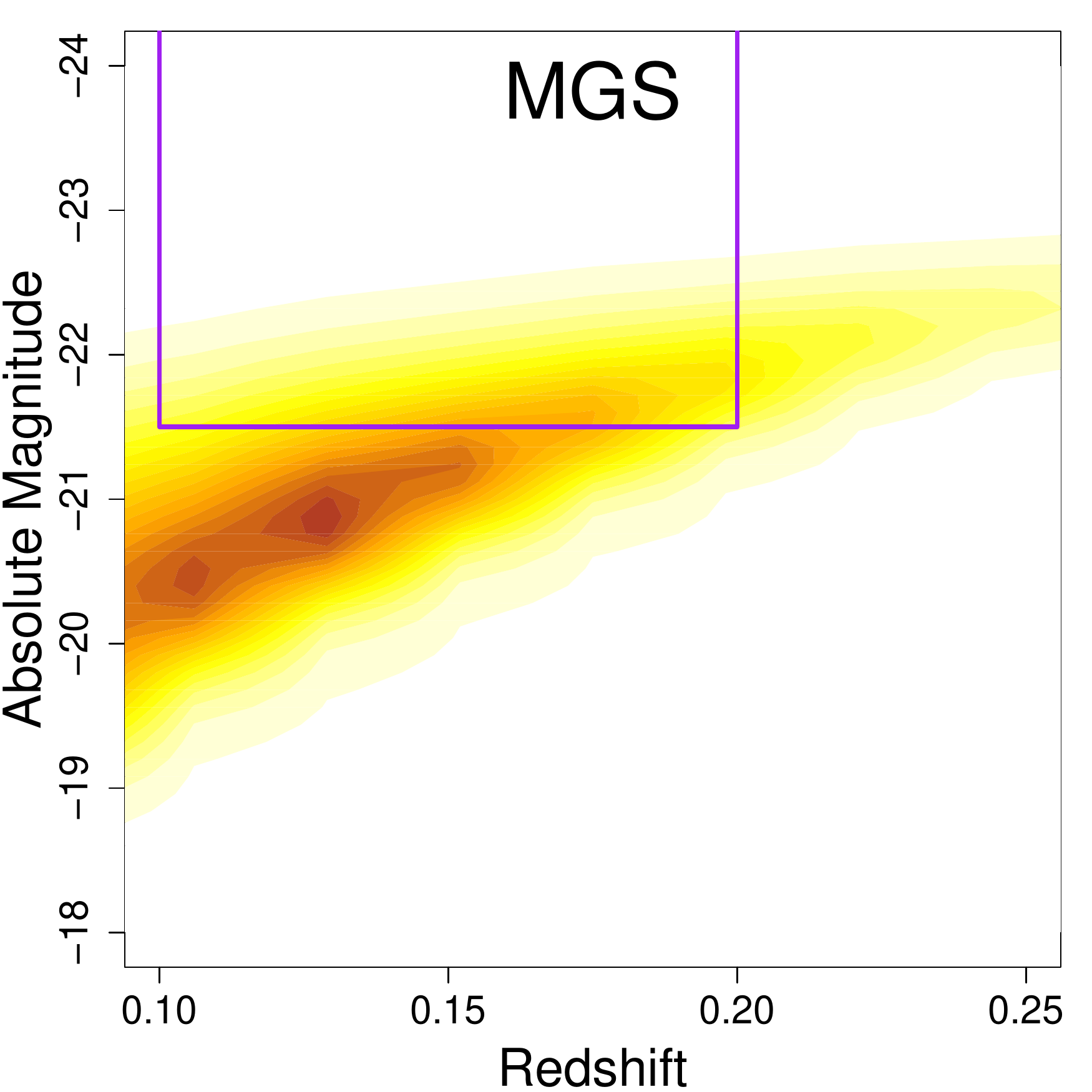}
	\includegraphics[height=2 in, width=2 in]{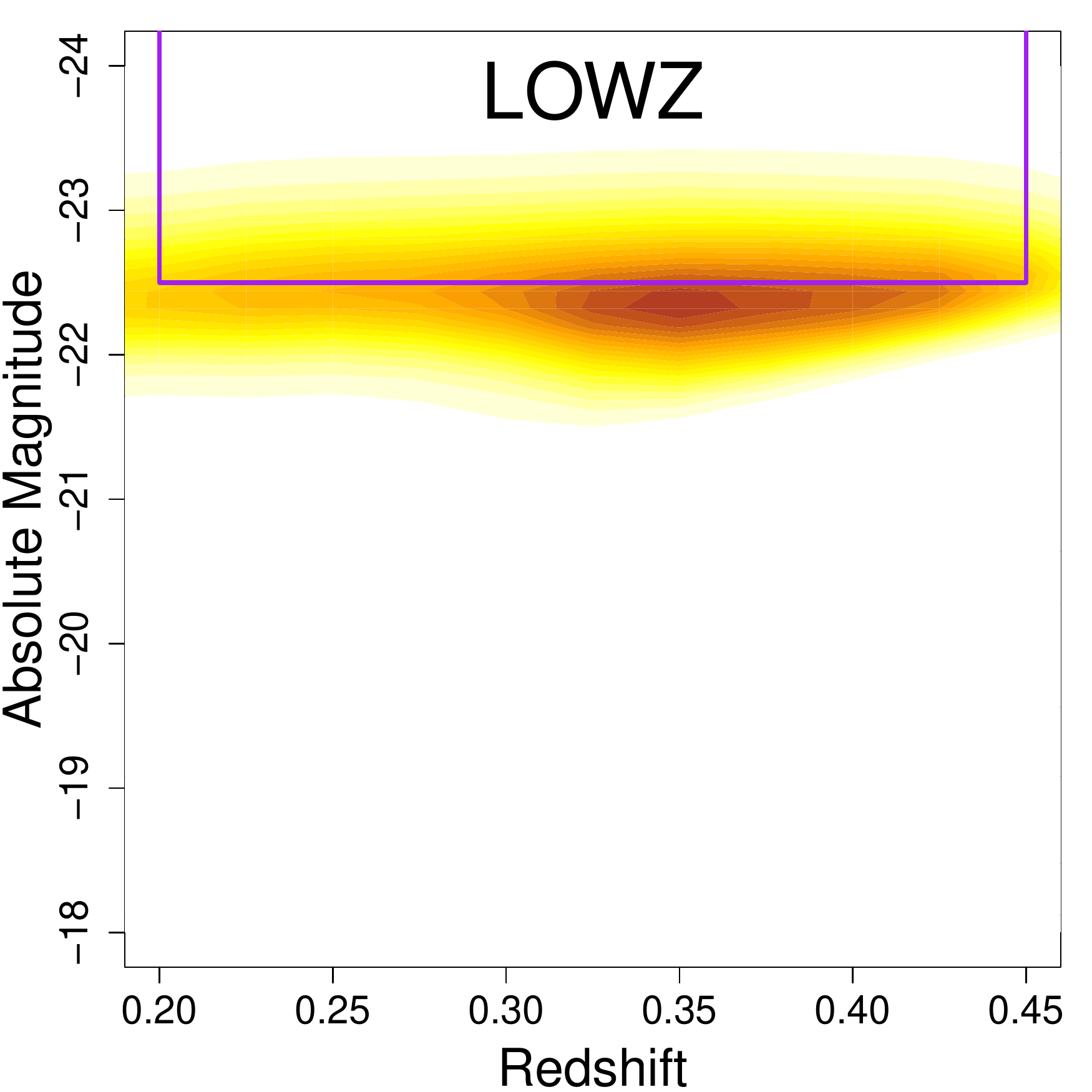}
	\includegraphics[height=2 in, width=2 in]{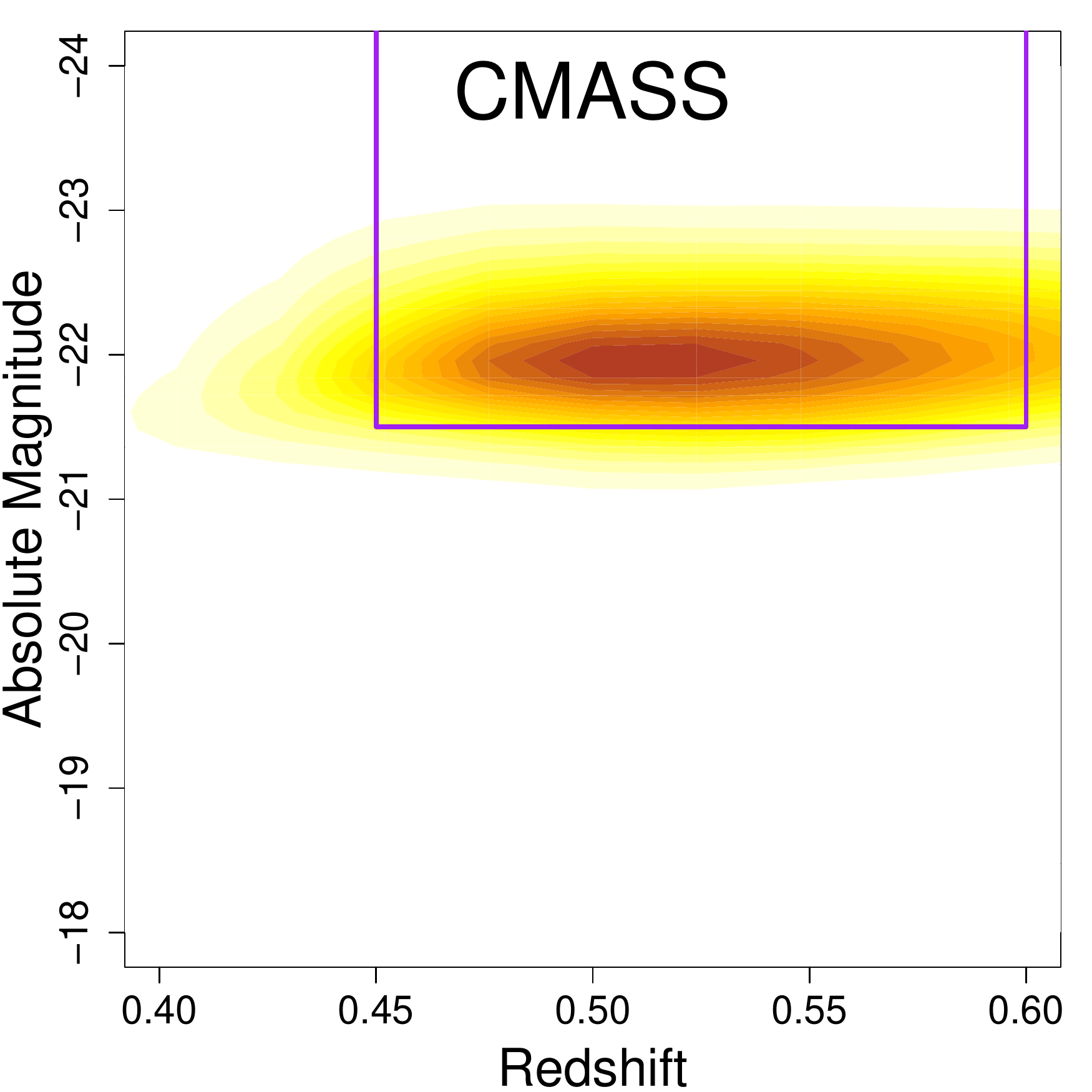}\\
	\includegraphics[height=2 in, width=2 in]{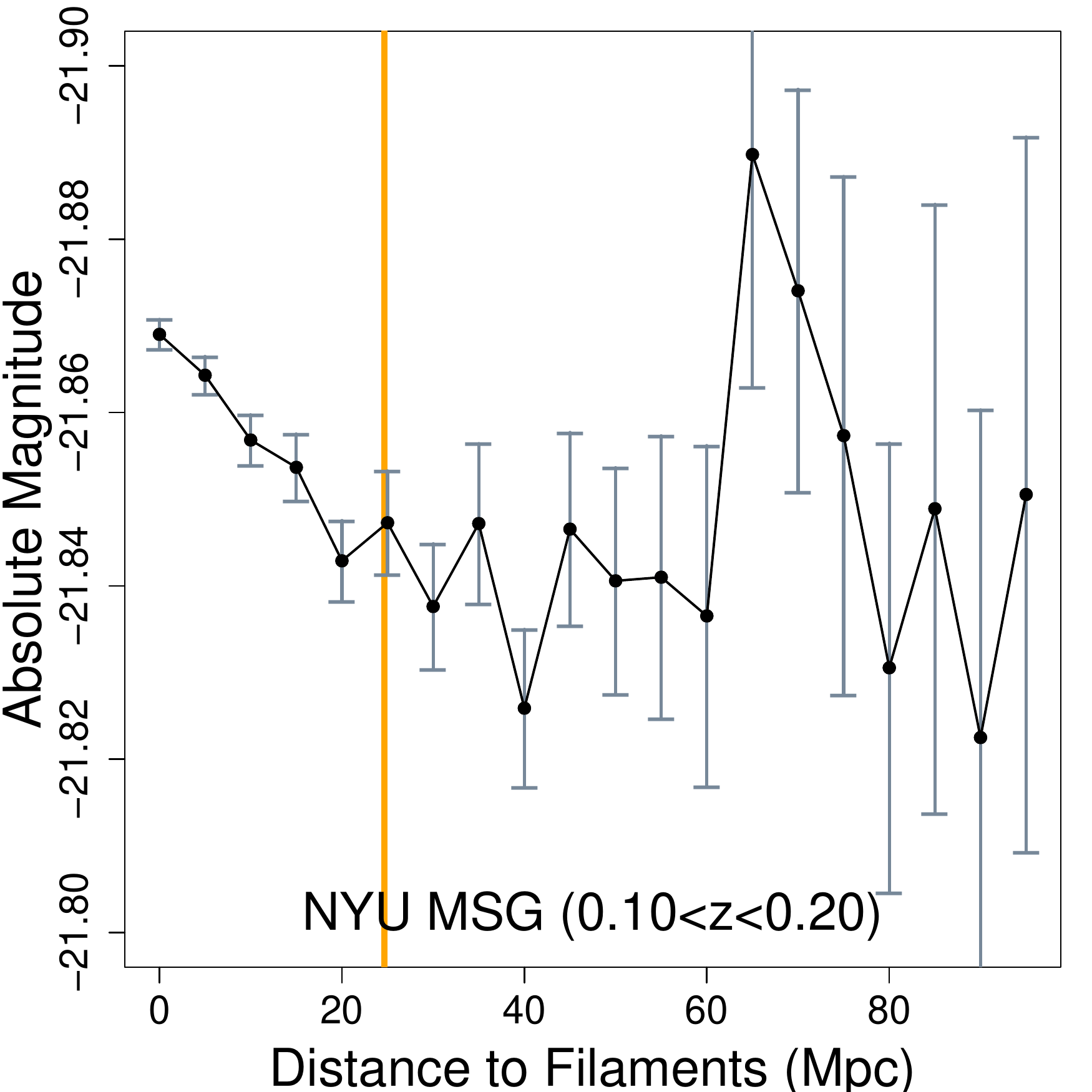}
	\includegraphics[height=2 in, width=2 in]{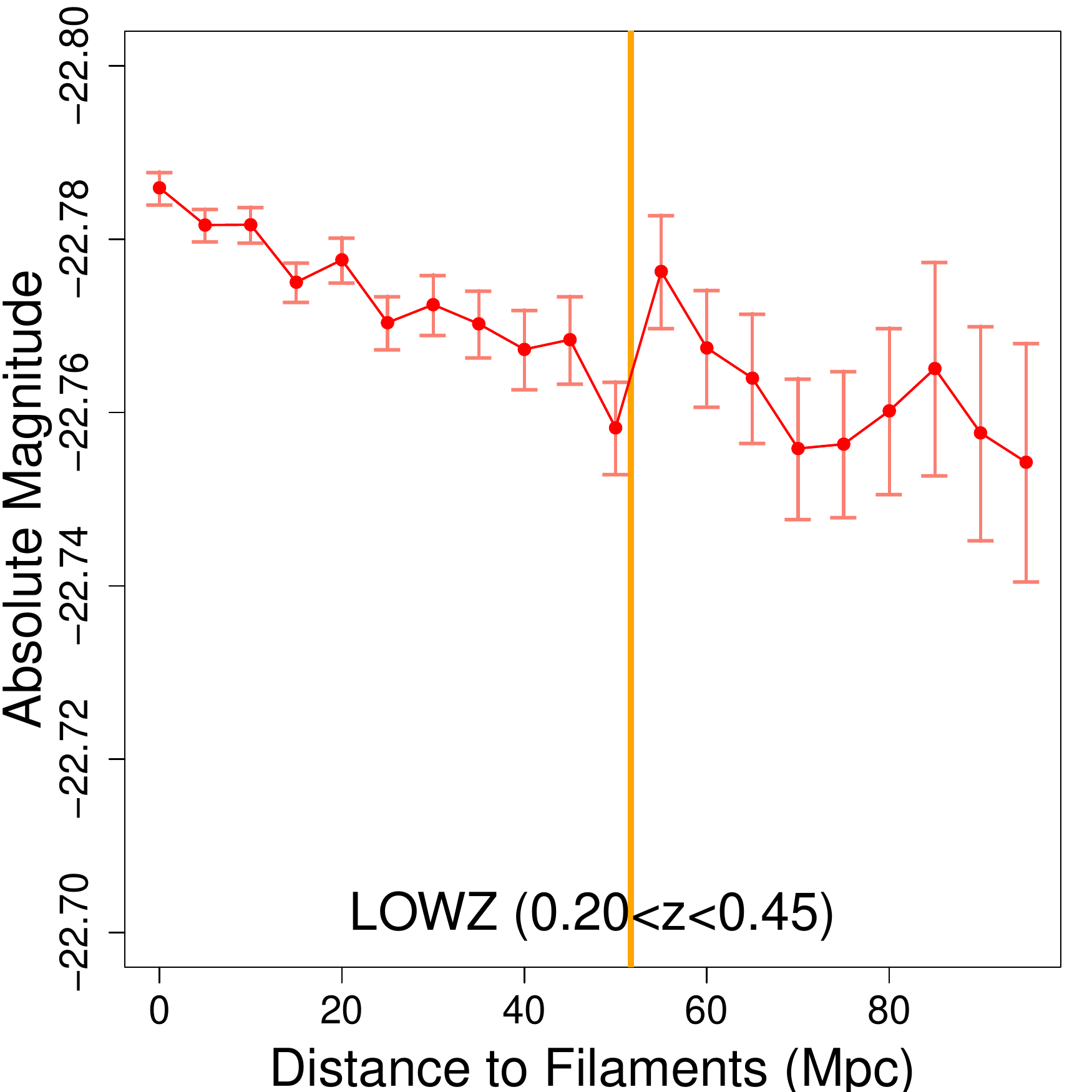}
	\includegraphics[height=2 in, width=2 in]{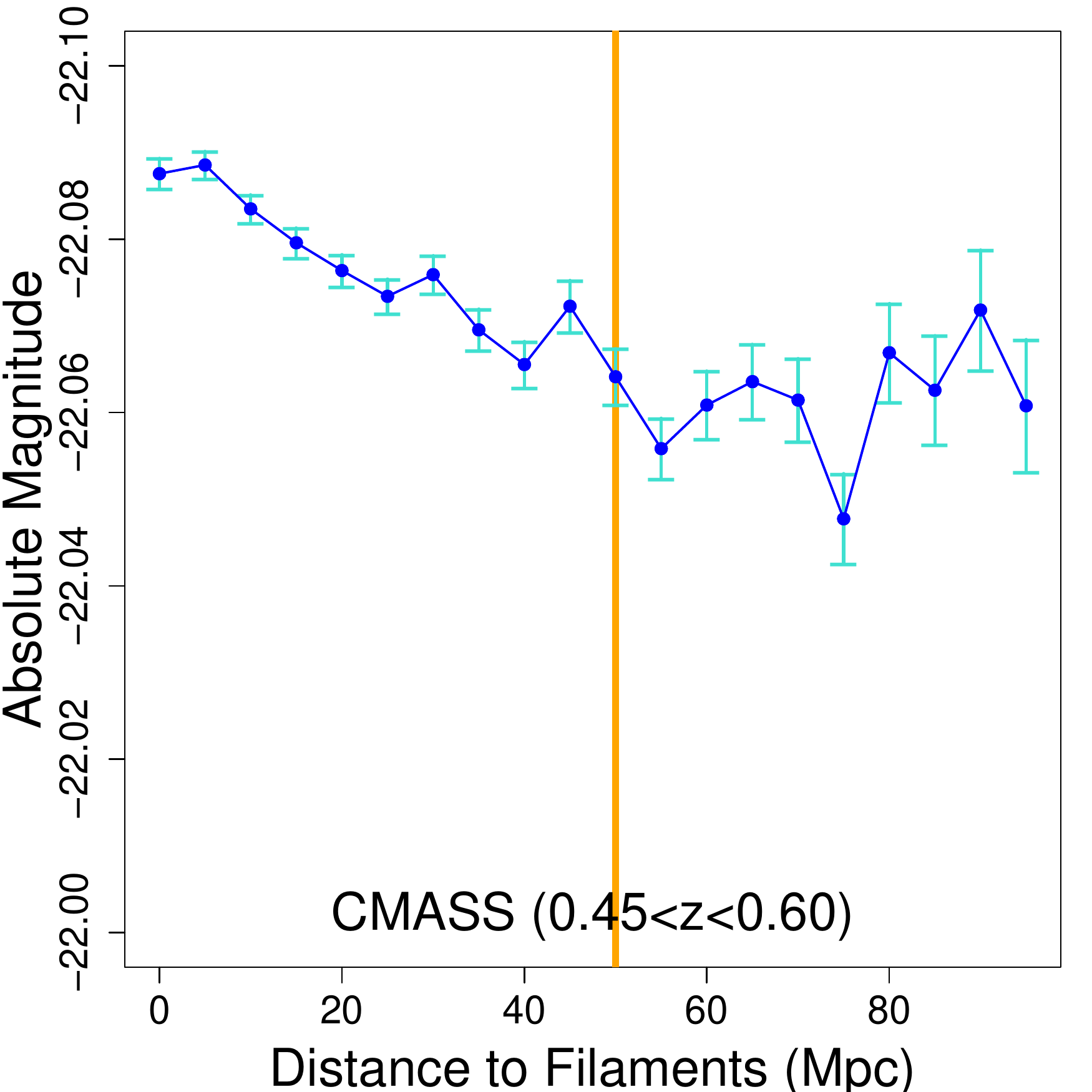}
\caption{
{\bf Top row}: Selected magnitude regions for each sample (purple rectangles).
Galaxies are selected within the purple rectangles to obtain
a volume-limited sample. 
There is a strong cut
on magnitude along the redshifts due to the observational limit.
We have reversed the direction for $Y$-axis (magnitude)
so that a galaxy in the upper region indicates that it is bright.
{\bf Bottom row}: Absolute magnitude ($r$-band) versus distance to filaments
for volume-limited samples. 
The orange line is the boundary between
decreasing pattern and random fluctuation.
A piecewise linear regression is fitted to the data to select the orange line
(selected values: $24.67$ Mpc for NYU MGS, $51.67$ Mpc for LOWZ and $57.34$
Mpc for CMASS).
On the left side of the orange line,
there is a strong decreasing trend, while on the right side,
the patterns exhibit random fluctuations.}
\label{fig::lum1}
\end{figure*}


We separately analyze each of the three galaxy catalogues
(NYU MGS, LOWZ and CMASS).
For each catalogue, 
we slice the redshift into several bins with width $\Delta z = 0.005$
that matches our filament maps.
We focus on the regions $0.1<z<0.6$ since the reMaPPer cluster catalogue
mainly covers this redshift range.
For each slice, we remove galaxies whose distance to galaxy clusters is less than $5$ Mpc,
thus eliminating the effect of galaxy clusters.

Since the SDSS dataset is not volume limited,
we had to apply additional constraints to construct a volume-limited sample.
Our selection rule is
\begin{align*}
(NYU\mbox{ }MGS)\quad &M_r < -21.5,\quad 0.10\leq z \leq 0.20\\
(LOWZ)\quad &M_r < -22.5,\quad 0.20\leq z \leq 0.45\\
(CMASS)\quad &M_r < -21.5,\quad 0.45\leq z \leq 0.60.
\end{align*}
The first row of Figure~\ref{fig::lum1} shows the luminosity-redshift region
within each of the NYU MGS, LOWZ and CMASS samples.
This figure reveals the strong luminosity-redshift dependence.


The bottom row of Figure~\ref{fig::lum1} presents the relation
between magnitude and distance to filaments for each sample.
Every sample processes a strong dependence of
magnitude and distance (to filaments). 
Galaxies near filaments are generally more luminous than
those at greater distance from filaments.
This relation
vanishes after certain range of distance (distance to the nearest filament).
To determine where the increasing pattern disappears,
we fit the following piecewise linear function:
\begin{equation}
M(x) = \left\{ 
  \begin{array}{l l}
    \beta_0 +\beta_1 x & \quad \text{if $x<x_c$}\\
    \beta_0 +\beta_1x_c & \quad \text{if $x\geq x_c$}
  \end{array} \right.
\label{eq::piece}
\end{equation}
where $M(x)$ is the magnitude and $x$ is the distance to filaments.
Namely, $M(x)$ is a linear curve when $x$ is less than the critical distance $x_c$
and is a constant after the critical distance.
The optimal fit suggests that $x_c$ for NYU MGS sample is $24.67$ Mpc,
for LOWZ sample is $51.67$ Mpc and for CMASS sample is $57.34$ Mpc.
This phenomena can be explained by the uncertainty of filaments.
The uncertainty in filaments will smooth out
the impact that the distance to filaments has on magnitude.
From Figure~\ref{fig::UM}, the uncertainties for filaments 
within the NYU MGS, LOWZ and CMASS samples are
$8, 15, 20$ Mpc, respectively. 
This is why the effect spans longer distances at high redshifts.

The slope $\beta_1$ in \eqref{eq::piece}
determines the strength, as well as the significance, 
for the decreasing pattern
and is given in Table \ref{tab::all-lum}.
According to Table \ref{tab::all-lum},
we observe a significant evidence (at $6.1\sigma-12.3\sigma$) 
that  the luminosity is indeed negatively correlated with the distance
before the critical distance.

\begin{table*}
\centering
\begin{tabular}{ccrrr}
  \hline
&$$ & NYU MGS ($z=0.10-0.20$) & LOWZ ($z=0.20-0.43$) & CMASS ($z=0.43-0.70$) \\ 
\hline
\multirow{3}{*}{Slope$^{\dagger}$}
&  Estimate & $-11.82\times10^{-04}$ & $-4.34\times 10^{-04}$ & $-5.13\times10^{-04}$ \\ 
&  Standard Error & $1.92\times10^{-04}$ & $6.53\times10^{-05} $& $4.17\times10^{-05}$  \\ 
&  Significance & $6.15 \sigma$ & $6.64\sigma$ &$ 12.31 \sigma$\\ 
   \hline
\end{tabular}
\caption{Linear fit for the three catalogues for absolute magnitude versus distance to filaments. 
$\dagger$ A negative slope indicates that the luminosity decreases as the distance (to filaments) increases.
}
\label{tab::all-lum}
\end{table*}

\section{Conclusion}
In this paper, we construct a series of two-dimensional filament maps
from SDSS data using the SCMS algorithm.
We provide several statistics to measure the properties of the filamentary 
maps we constructed at each redshift.
These measurements may be used
to study the evolution of the Universe and constrain cosmology.

We compare our publicly available catalogue to 
the existing catalogues for filaments introduced in \cite{2008ApJ...672L...1S}, 
\cite{2010MNRAS.409..355J}, \cite{2012MNRAS.422...25S}, and \cite{2014MNRAS.438.3465T}.
Each of these catalogues provide some analysis for the large-scale structure
over the whole Universe by using different models for filaments.
However, none of them is publicly available.
This makes it difficult for other research groups to use these catalogues
to analyze filaments.
Moreover, unlike our catalogue all these catalogues do not provide any measurement on the errors
for filament detection and only focus on the small redshift range (less than $z=0.25$).
To our knowledge, our filament catalogue is by far the only filament catalogue
for redshift $z>0.25$ in the SDSS.


We apply our filament maps to investigate the galaxy luminosity-filament distance relarion
using a volume-limited sample.
There is a long distance effect
from filaments (more than $20$ Mpc) on the brightness of galaxies, 
which is at a different scale than
\cite{2015ApJ...800..112G}, where they found a similar pattern
at a much smaller scale (distances less than $0.71$ Mpc).
Although part of the long distance effect can be explained by
the errors of filaments, our results suggest that
the correlation between galaxy magnitude and distance to filaments may extend
over distances $>> 1$ Mpc.


\section*{Acknowledgments}
We thank Hung-Jin Huang, Rachel Mandelbaum, Michael Strauss, and Hy Trac for useful discussions
and comments. 
This work is supported in part by the Department of Energy under grant DESC0011114; 
YC is supported by William S. Dietrich II Presidential Ph.D. Fellowship;
SH is supported in part by DOE-ASC, NASA and NSF; CG
is supported in part by DOE and NSF;
LW is supported by NSF.
Funding for SDSS-III has been provided by the Alfred P. Sloan Foundation, the Participating Institutions, the National Science Foundation, and the U.S. Department of Energy Office of Science. The SDSS-III web site is http://www.sdss3.org/.

SDSS-III is managed by the Astrophysical Research Consortium for the Participating Institutions of the SDSS-III Collaboration including the University of Arizona, the Brazilian Participation Group, Brookhaven National Laboratory, Carnegie Mellon University, University of Florida, the French Participation Group, the German Participation Group, Harvard University, the Instituto de Astrofisica de Canarias, the Michigan State/Notre Dame/JINA Participation Group, Johns Hopkins University, Lawrence Berkeley National Laboratory, Max Planck Institute for Astrophysics, Max Planck Institute for Extraterrestrial Physics, New Mexico State University, New York University, Ohio State University, Pennsylvania State University, University of Portsmouth, Princeton University, the Spanish Participation Group, University of Tokyo, University of Utah, Vanderbilt University, University of Virginia, University of Washington, and Yale University.

\appendix

\section{Algorithm for Detecting Intersection Points}
\label{sec::int::alg}
In this section, we describe the metric graph reconstruction algorithm
\citep{Aanjaneya2012, 2013arXiv1305.1212L} for detecting
intersection points of filaments.
Our algorithm examines every point on the filaments
and assigns it into the
`intersection' class or `non-intersection' class
using the following process.
Let $x$ be a point we wish to examine.
\begin{itemize}
\item[1.] Keep those data points whose distance to $x$ is between 
$r_{in}$ and $r_{out}$, two parameters.
\item[2.] Cluster the remaining points using hierarchical clustering 
with radius $r_{sep}$, i.e., partitioning points into several groups
such that group-group distance is greater than $r_{sep}$.
\item[3.] Count the number of groups from previous step. 
If the number of groups is greater or equal to three, 
classify $x$ as an \emph{intersection} point, otherwise classify it as \emph{non-intersection}.
\end{itemize}
The idea behind this algorithm is that 
when a point is at the intersection,
other points around this point within the shell (form by $r_{in}$ and $r_{out}$)
should have at least three clusters.
For an edge point, there will be two clusters and 
for the end point, there is only one cluster.
Points near the same intersection may all be classified
as intersection points; we use the mean location as
intersection point:
\begin{equation}
r_{in}= 2h/3, \quad r_{out} = 2 r_{in}, \quad r_{sep} = (r_{in}+r_{out})/2.
\end{equation}
This choice of parameters is ad hoc but works well in practice.

\bibliographystyle{mnras}
\bibliography{SuRF_cata.bib}
\bsp

\label{lastpage}

\end{document}